\documentstyle[11pt,epsfig]{report}

\def\be{\begin{equation}}
\def\ee{\end{equation}}
\def\bea{\begin{eqnarray}}
\def\eea{\end{eqnarray}}
\def\gm{\Gamma}
\def\bt{\begin{table}}
\def\et{\end{table}}
\def\ra{\rightarrow}
\def\ln{\left<}
\def\rn{\right>}
\def\lln{\left(}
\def\rln{\right)}
\def\lq{\Lambda_{QCD}}
\def\bl{\bar \Lambda}

\def\dl{\bar \Lambda - \delta m}
\def\tb{\tau}
\def\lbb{\Lambda_b}
\def\lq{\Lambda_{QCD}}
\def\bl{\bar \Lambda}

\def\dl{\bar \Lambda - \delta m}
\def\ra{\rightarrow}

\def\om{\omega}
\def\qq{\ln \bar q q \rn}
\def\al{\alpha}

\def\np{Nucl. Phys.~}
\def\pl{Phys. Lett.~}
\def\prl{Phys. Rev. Lett.~}
\def\pr{Phys. Rev.~}
\def\prr{Phys. Reports~}

\def\ep{Eur. Phys. J~}
\def\bb{\bibitem}

\begin{document}
\textwidth17.0cm
\textheight20.0cm
\baselineskip = 1.2\baselineskip
\begin{titlepage}
\begin{center}
{\bf SPECTATOR EFFECTS AND \\
QUARK-HADRON DUALITY IN\\
INCLUSIVE BEAUTY DECAYS}
\vskip0.75cm

\vskip0.25cm
{\bf S. ARUNAGIRI, M.Sc.,}
\vskip4.75cm
Department of Nuclear Physics\\
School of Physical Sciences\\
University of Madras\\
Guindy Campus\\
Chennai 600 025, INDIA
\vskip0.7cm

\end{center}
\end{titlepage}

\pagenumbering{roman}
\tableofcontents
\newpage
\section{Preface}

This Thesis is a review of the inclusive decays of heavy hadron. It is based on
the author's study of
two issues which are important in the description
of the inclusive beauty decays given by the heavy quark expansion.

\begin{itemize}
\item
Evaluation of the expectation values of
the four-quark operators in the heavy quark
expansion,

\item
Study on the validity of the
quark-hadron duality in the heavy quark
expansion. 
\end{itemize}

We evaluate the four-quark operators in two approaches:
(1) in terms of the oscillator strength of the hadrons
in the harmonic oscillator model, and (2) from the difference in the
total decay rates of beauty hadrons under $SU(3)$ flavour symmetry.
The results obtained indicate that the lifetime difference between
$B$ meson and $\Lambda_b$ can be explained within heavy quark expansion
by including the third order contribution.

To discuss the quark-hadron duality, the power corrections to the
partonic decays rate are obtained in terms of the gluon mass
which is assumed to imitate the short distance nonperturbative effects,
by the choice $\lambda^2 \gg \Lambda_{QCD}^2$.

The Thesis is organised in six Chapters, as follows:
\begin{itemize}
\item
{\bf 1 Introduction:} The Standard Model is briefly discussed and then the
Spin-flavour symmetry exhibited by heavy hadrons is reviewed.

\item
{\bf 2 Four-quark operators:} The author's works on the
evaluation of four-quark operators are presented.

\item
{\bf 3 Spectator effects:} Presented are the
calculation of ratio of lifetimes of beauty hadrons,
particularly, $\tau(\Lambda_b)/\tau(B)$ and the inclusive
charmless semileptonic decay rate and determination of the CKM matrix element
$|V_{ub}|$.

\item
{\bf 4 Short distance nonperturbative corrections:} The renormalons contribution
to the heavy quark expansion is discussed based on the authors work.

\item
{\bf 5 Quark-hadron duality:} The validity of the assumption
of quark-hadron duality is qualitatively discussed.

\item
{\bf 6 Conclusion:}
Conclusion of the Thesis with outlook of the works
done are given.

I am glad to acknowledge that my acquaintance with heavy quark physics
began with the lectures on the heavy quark symmetry by
{\bf Prof. Apoorva Patel}
during a discussion meeting at the Centre for Theoretical Studies,
Indian Institute of Science, Bangalore. I got this great opportunity
due to the invitation of {\bf Prof. J. Pasupathy}. My learning of QCD is
due to them by the way of their invitation to the discussion meetings
conducted by them. I am grateful to both
of them for their continuing support.

\end{itemize}

\vskip2.0cm
\begin{flushright}
S. ARUNAGIRI
\end{flushright}

\newpage

\section{Acknowledgements}

I thank Prof. P. R. Subramanian, Head of this Department, 
who has supervised my Thesis work and encouraged me to do things independently.

I am greatly indebted to Prof. H. S. Sarath Chandra, The Institute of Mathematical 
Sciences, Chennai, for discussions
and kind attitude towards me.
I place on record my deep gratitude to Prof. Hitoshi Yamamoto, University
of Hawaii, USA, for a collaboration and sharing his wisdom. My sincere 
gratitude are due to Prof. Yasuhiro Okada, Theory Group, KEK, Japan,
for extending me an invitation to KEK and useful discussions. I thank
Prof. S. Narison, University of Montpellier, France, who
clarified many doubts on renormalons.

I am grateful to Prof. M. Neubert, Cornell University for useful
communications. My sincere thanks are due to Prof. M. Voloshin, University
of Minnesota, USA for clarifications and discussions.

I enjoyed the comradeship with 
Dr. D. Caleb Chanthi Raj,
SRM Engineering College, Chennai, Dr. J. Segar, RKM Vivekananda College, Chennai,
and Dr. R. Premanand, SRM Engineering College, Chennai. With their helps
and affection only, I derived strength to bear joys and sorrows during this 
programme.

I acknowledge the interests shown on me and my works by
Prof. V. Devanathan and Prof. T. Nagarajan, former heads of
this Department, the faculty members and other staff
of the Department of Nuclear Physics.

It is a pleasure for me to be with Prof. K. Raja,
Dr. N. Baskaran, Dr. Manivel Raja, Dr. R. N. Viswanath, Dr. R. Ramamoorthy,
Messers A. Chandra Bose, N. Ponpandian, R. Justin Joseyphus,
K. Ravichandran, R. Shanmugam, B. Nagarajan,
B. Soundarajan and R. Sudharsan. I thank them
for their help.

I thank the authorities of The Institute of Mathematical Sciences
for allowing me to use their library and computing facilities. 

I am grateful to
Mr. V. Swaminathan who has been pouring passionate interest in me and 
my scheme of things and Mr. C. Krishnamoorthy for his helps.

I express my deep sense of gratitude to (Late) Mr M. Arangarajan 
who had inspired me as a brother, a mentor and a friend from my childhood
and his family members. I appreciate the forbearance of my father 
Mr R. Somasundaram and the other members of
our family who stood by me in successful completion of this programme.

\vskip2.0cm
\begin{flushright}
S. ARUNAGIRI
\end{flushright}

\newpage

\section{List of Publications}
\begin{enumerate}

\item
{Colour-straight four-quark operators and lifetimes of b-flavoured hadrons,\\
S. Arunagiri, {\it Int. J. Mod. Phys.} {\bf A 15} (2000) 3053 - 3063 [hep-ph/9903293]}

\item
{A note on spectator effects and quark-hadron duality in inclusive
beauty decays\\
S. Arunagiri, {\it Phys. Lett.} {\bf B 489} (2000) 309-312 
[hep-ph/0002295].}

\item
{Inclusive charmless semileptonic decay of $\Lambda_b$
and $Br(b \ra X_u l \nu)$\\
S. Arunagiri and H. Yamamoto [hep-ph/0005266].}

\item
{On the  short distance nonperturbative
corrections in heavy quark expansion\\
S. Arunagiri, presented in the XVIII International Symposium on
Lattice Field Theory, held at Bangalore, India, during August 17-22,
2000 [hep-ph/0009109].}

\item
{Spectator effects in inclusive beauty decays\\
Contributed to the Working group report on B and collider physics, 
Proc. VI Workshop on High Energy Physics Phenomenology,
held at Chennai, India, during January 3-15, 2000,\\
Pramana - J. Phys. {\bf 55} (2000) 335-345 [hep-ph/0004002].}

\item
{Pattern of lifetimes of beauty hadrons and quark-hadron duality
in heavy quark expansion\\
S. Arunagiri, to be published in the Proceedings of IV Workshop on
Continuos Advances in QCD, held at Minneapolis, USA, during May 12 - 14,
2000 [hep-ph/0009110]. }

\end{enumerate}

\chapter{Introduction}
\pagenumbering{arabic}
This Chapter consists of two sections: The first is devoted to the Standard Model  
and in the second the heavy quark symmetry  and the  heavy quark expansion are discussed
\footnote{Natural units, namely, $\hbar = c = 1$, are used throughout.}. 
\section{Standard model}

In this section, we briefly describe the texture of the Standard Model
and remark on the perspectives of the weak decays of heavy hadrons in
the Standard Model. 
\subsection{Quarks, Leptons, and Hadrons}

Today, our knowledge is, as embodied in the Standard Model, firm about the primitive 
constituents of matter 
\cite{halz}. They
are six quarks and six leptons, all fermions. The quarks are interacting
through all the known three interactions of the elementary particles in nature:
electromagnetic, weak  and strong, but the leptons do not {\it strongly} interact.
Both of them are not influenced by the gravity effects. The interactions 
are mediated by (gauge) bosons: photon, $W^\pm, Z^0$ and gluon. 
The properties of these fermions and bosons are given in Tables \ref{t1} 
and \ref{t2} \cite{pdg}. 

The fermions make up the hadrons: meson, a bosonic hadron, of a quark-antiquark
pair and baryon, a fermionic hadron, of three quarks. Quark structure of 
hadrons is described, for three quarks $u, d$ and $s$, under $SU(3)$ flavour symmetry
by the 
Gell-Mann - Nishijima relation:
\be
Q = I_z + (B+S)/2 \label{gm}
\ee
where $Q$ is the electric charge of the hadron, in units of $e$, $I_z$ the
third component of the isospin, $B$ the baryon number, and $S$ the strangeness
quantum number. Each quark is assigned a colour quantum number. Each quark 
carries three colours but the quarks make hadrons colour singlet.
For mesons, $B = 0$. This can be extended to the
charm and beauty quarks as well. The situation is different in the case of top which decays before
hadronising. The rate of top quark weak decay, 
$\Gamma (t \ra b W)$ \cite{xxx}, is found to be larger
than the QCD scale that corresponds to the typical hadron size. 
 
\bt
\caption{Quarks and Leptons ($^a$ in GeV/c$^2$; $^b$ in $e$ units; $^c$ 
quantum numbers; $^d$ charge of leptons except neutrinos
is $-1$)}
\begin{tabular}{|c|c|c|c||c|c|c|} \hline

{\bf Quark}		& Mass$^a$ & 
Charge$^b$ & Qu. nos.$^c$ &
{\bf Lepton}$^d$		& Mass	 \\ \hline

up, $u$		& (1.5 - 5) $.10^{-3}$  & $ +{2 \over 3}$  & 
$I_z = + {1 \over 2}$ &
$e^-$ & 0.51 $.10^{-3}$	\\

down, $d$	& (3 - 9) $.10^{-3}$    & $-{1 \over 3}$ & 
$I_z = - {1 \over 2}$ &
$\mu^-$   & 0.105		 \\

strange, $s$	& 0.060 - 0.170  & $-{1 \over 3}$ & $s = - 1 $   &
$\tau^-$   & 1.777		 \\
 
charm, $c$	& 1.1 - 1.4 & $+{2 \over 3}$  & $c = 1 $ &
$\nu_e$		& 10-15 $.10^{-6}$ \\

bottom, $b$	& 4.1 - 4.4 & $-{1 \over 3}$ & $b = -1 $ &
$\nu_\mu$	& $ < 0.17 . 10^{-3} $ 	\\

top, $t$	& 170       & $+{2 \over 3}$  & $t = 1 $  &
$\nu_\tau$	& $ < 18 .10^{-3}$ \\ \hline
\end{tabular}
\label{t1}
\et

\bt
\caption{Force carriers}
\begin{center}
\begin{tabular}{|c|c|c|c|c|} \hline
Gauge Boson	& Mass 	& Charge 	&  Force & Strength\\ \hline
photon, $\gamma$& 0		& 0		& electromagnetic & 1/137 \\
$W$    		& 80		& $\pm$ 1	& weak & 1.1663692 GeV$^{-2}$\\
$Z$		& 90		& 0		& weak &\\
gluon, $g$	& 0		& 0		& strong & 0.12 at $M_Z$\\ \hline
\end{tabular}
\label{t2}
\end{center}
\et

In the next section we present a brief review of the Standard Model (SM).
The SM is a renormalisable field theory of the local gauge
group $SU(3)_C \times SU(2)_L \times U(1)_Y$ \cite{sm}. It is a unified theory
of strong ($SU(3)_C$), weak ($SU(2)_L$) and electromagnetic ($U(1)_Y$)
interactions at $TeV$ scale. The SM contains two sectors: quantum
chromodynamics and electroweak \cite{don}.

\subsection{Quantum Chromodynamics }

\subsubsection{A prelude}
In the proposal of quark structure of hadrons made by Gell-Mann and Zweig, 
the quarks are described in terms of flavour 
and colour as required to explain the known properties of 
hadrons.  The idea of quarks carrying colour got confirmed
by the measured normalisation of the decay rate $\Gamma (\pi^0  \ra 2\gamma)$
and the cross-section $\sigma(e^+e^- \ra hadrons)$. Using current algebra,
the strong interaction between quarks is shown to be mediated by 
massless vector bosons, gluons, analogous to photon of electromagnetic 
interaction. According to the colour assignment to quarks, 
gluons form an octet.

Lepton ($e^-/\nu$) - proton deep inelastic scattering experiments confirmed 
the composite structure of a proton. Further, from the deep inelastic scattering
experiments, it was found that quarks and gluons are
freely roaming around {\it inside} the proton. This behaviour
is explained by Bjorken through his scaling law that at small distance 
(this corresponds to large momentum transferred),
the quarks are behaving as if they are free, but at large distance ({\it i. e.,}
small momentum transferred), they are confined in pair. The former
property is known as {\it asymptotic freedom} and the latter {\it
confinement}.  

\subsubsection{The Lagrangian of QCD}
Theory of quarks and gluons is based on the local non-abelian gauge group $SU(3)_C$
symmetry\cite{ynd}. With the work of 't Hooft on renormalisability of Yang-Mills theories \cite{hooft},
the theory of strong interaction is formulated. The basic Lagrangian is
\be
{\cal L}_{QCD} = -{1 \over 4}F^a_{\mu \nu}F_a^{\mu \nu} + 
	\sum_f \bar q^{f,a}(x)i\gamma_\mu D^\mu q(x)^f_b -
	\sum_f m_f \bar q^{f,a}(x)q(x)^f_b 
\label{l1}
\ee
where the indices $a, b$ stand for colour and $f$ for flavour of quarks and
\bea
F_{\mu \nu}^a &=& \partial_\mu A_\nu^a-\partial_\nu A_\mu^a - 
	g_s f^{abc} A_\mu^b A_\nu^c, \hspace{0.5cm} \\
D_\mu^a &=& \partial_\mu^a+ig_st^aA_\mu^a
\eea
where $g_s$ is the strong coupling constant, 
$f^{abc}$ the structure constant of $SU(3)$ Lie algebra and $t^a = 
{1 \over 2} \lambda^a$, with $\lambda^a$ being the 
Gell-Mann colour $(3 \times 3)$ matrices. In Eq.
(\ref{l1}), the first term is the gluon part, the second the quarks
part and the quark-gluon interaction part and the third the quark mass 
contribution.
The Lagrangian is invariant under the local gauge group $SU(3)$. 
The gauge transformation is given by
$U(x) = e^{it^a \xi^a(x)}$
where $\xi^a(x)$ is an infinitesimal gauge parameter with $a$ running from 
1 to 8. The $t^a$ are the generators of the gauge group.
Because of the non-Abelian nature of the boson field, 
gluons can be described as self-interacting, vector
bosons. The Lagrangian, in (\ref{l1}), is supplemented by 
additional terms corresponding to gauge-fixing, ghost and counter terms
for perturbative treatment. With these terms, one can get the Feynman rules
for QCD.

The QCD Lagrangian, in the limit of $m_q = 0$, where $q = u, d, s$, is invariant under
the chiral symmetry group $SU(3)_L \times SU(3)_R$ \cite{luw}. On the other extreme, for
quarks of mass $m_Q \gg \Lambda_{QCD}$, in the limit $m_Q \ra \infty$, the 
Lagrangian exhibits spin-flavour symmetry \cite{neub} which will be discussed in details
in the next chapter. Besides , ${\cal L}_{QCD}$ possesses
stability against weak radiative corrections due to the flavour-neutrality
of the gluon. 

\subsubsection{Perturbative QCD: Running Coupling and Running Mass}
Different renormalisation schemes are used to make the Lagrangian 
free of divergences. Then the renormalised quantities are defined at 
some chosen scale \cite{paus}. Different choice of the scale leads to
different physical values for the same quantity. In order to circumvent
the scale dependence, renormalisation group equations are used. 
The scale transformation,
$\mu \ra \mu^\prime$, constitutes a renormalisation group. Under this group,
the physical quantities, explicitly or implicitly scale dependent,
are invariant. Such a study is due to Callan-Symansik.
In the $MS$ or $\overline {MS}$ scheme, for the coupling constant, the 
Callan-Symasik or renormalisation group equation is given by
\be
d \bar g(\mu)/d\log \mu = \bar g(\mu) \beta(\bar g(\mu))
\ee
An expansion of $\beta$ in $\bar g(\mu)$ yields, at the lowest order,
\be
\beta = - \beta_0 {\bar g(\mu)^2 \over {16 \pi^2}}, \hspace{0.5cm}
\beta_0 = {11 C_A \over 3} - {4 n_f T_F \over 3} 
\ee
where $C_A = 3, T_F = 1/2$ and $n_f$ is the number of flavours. Now,
defining the running coupling constant $\alpha_s = \bar g^2/4\pi$, 
with $\mu = Q^2/\nu^2$, 
\bea
\alpha(Q^2) = {4\pi \over {\beta_0 \log\left({Q^2/{\Lambda^2}}\right)}},\\
\Lambda^2 = \nu^2 e^{-{4\pi /{\alpha(\nu^2)}}} \label{alp}
\eea
where $\lq$ is known as the QCD scale parameter, which is 
the constant of integration. From (\ref{alp}), it is obvious that 
as $Q^2 \ra \infty, \alpha_s \ra 0$. This behaviour is the celebrated
asymptotic freedom \cite{gross}. For finite but sufficiently
large $Q^2$, $\alpha_s$ is
employable as a perturbation parameter.

Similarly, the running mass $\bar m(Q^2)$ is given by
\be
\bar m(Q^2) = m_0 \left[{1 \over 2}\log \left({Q^2 \over {\Lambda^2}}\right)\right]^{-d_m} 
\label{m}
\ee
where $m_0$ is the invariant mass and $d_m$ is known as the
anomalous mass dimension, given by $4/\beta_0$.

\subsubsection{\it A digression}
The perturbation series of a physical quantity in terms of $\alpha_s$ is known
to few orders. A quantity is said to be defined completely, if the expansion
in $\alpha_s$ is made to {\it all} orders. But it
is impossible to define them to $n$ orders, where $n$ is large. At large order,
as first discovered in QED by Dyson in 1952, the series in $\alpha_s$ 
factorially diverges. This issue has attracted interests 
recently after it was first pointed out by 't Hooft in 1977 that the inherent 
ambiguities in the perturbative definition of a quantity
is $O\left({\Lambda_{QCD}/Q}\right)$ and beyond. These ambiguities
are known as renormalons \cite{zak}. These corrections are seen to be important
as they would represent the first nonperturbative corrections to the
leading order of the OPE.

The QCD needs to be dealt with at small and large distance scales separately. 
Any physical amplitude of a QCD process described by the product of operators
can be separated into two parts by the Wilson short distance or operator
product expansion \cite{wil}. As a result, the short distance physics is contained
in the expansion coefficients, which happen to be $c$-numbers, and the local 
operators describe the long distance aspect. 
While the perturbative QCD successfully describes the short distance processes, 
the long distance aspects being nonperturbative are poorly understood. However,
there are semi analytic methods 
such as QCD or SVZ sum rules \cite{shif0} and lattice gauge theory \cite{cru}.

\subsection{Electroweak Theory}
\subsubsection{Fermi theory of weak interaction}
With the explanation of the the radioactive decay of a nucleus, the weak
force was discovered. For example, 
weak interaction drives a neutron to change into a proton
with a lepton pair, electron and its antineutrino, inside the nucleus.
At the quark level, it is the transition of the $d$ quark into the $u$ quark plus
a lepton pair. A 
theoretical description of weak interaction emerged due to  
Yang and Lee, and Wu et al \cite{fermi} following Fermi. The Lagrangian of weak interaction is then given
by
\be
{\cal L}_{Fermi} = {G_f \over {\sqrt{2}}}J^\mu(x)J_\mu^{\dag} (x) + h. c.,
\label{lf}
\ee
where $G_f = 1.16637 \pm 0.00001 GeV^{-2}$ is the Fermi coupling constant 
and the weak current $J^\mu$ is given by
\be
J^\mu(x) = \bar p(x) \gamma^\mu (g_V-g_A \gamma_5) n(x) \bar e^-(x) 
\gamma_\mu (1-\gamma_5) \nu_{e^-}(x),   
\ee
where the vector coupling $g_V$ is little less than one and is given in terms of the
Cabibbo angle to be 0.97. The ratio of the coupling constants $g_A/g_V = -1.2573 
\pm 0.0028$ \cite{pdg}. 

Towards formulating a complete theory of weak interaction, it is postulated that
the weak interaction is mediated by a vector boson, massive due to weak 
interaction being point-like. The additional requirements of such a
boson are that it has to be charged both positively and negatively because
of charge-changing weak current and having indefinite parity to preserve the
$V-A$ structure of the weak current. Thus, the Lagrangian is 
\be
{\cal L}_W = -g_W J^\mu(x)W_\mu(x) + h. c.
\label{lw}
\ee
where $W(x)$ is the vector boson field and $g_W$ the dimensionless weak coupling
constant which is related to the Fermi coupling constant by
\be
{g_W^2 \over {8M_W^2}} = {G_f \over \sqrt{2}}
\ee
The Lagrangian in (\ref{lw}) is useful at low energy but not renormalisable. 
The existence of $W^\pm$
is another way of looking at $G_F$. The step towards the renormalisable
theory of weak interaction results in the unification of the weak and the 
electromagnetic interactions, {\it electroweak interaction}.

\subsubsection{An interlude}
The current-current interaction described by (\ref{lw}) is applicable only for
charged current. There is also a neutral current mediated weak interaction. If
both are combined, then three massive vector bosons correspond to the isospin
triplet of weak currents. This obviously leads to $SU(2)_L$ symmetry of 
weak interaction.  The fermion field is a left handed doublet. Then, the current
is of the form
\be
J_\mu^i(x) = (\bar \nu \hspace{0.35cm} \bar e)_L\gamma_\mu {\tau_i \over 2}   
\left(
\begin{array}{l}
\nu \\ 
e^-
\end{array}
 \right)_L
\label{J}
\ee
where $\tau_i$ is the isospin triplet. Since the weak neutral current,
$J_\mu^{NC}$, has a right handed component, the neutral component of (\ref{J})
is not a weak current. However, this can be identified with the electromagnetic
current
\be
j_\mu = ej_\mu^{em} = e\bar \Psi \gamma_\mu Q \Psi
\ee
where $Q$ is the charge operator with eigenvalue --1 for electron and the
generator of $U(1)$ symmetry. Under orthogonal combination
\be
j_\mu^{em} = J_\mu^3 + {1 \over 2}J_\mu^Y
\ee
The consequence of the above equation is the relation, $Q = T^3_W + Y_W/2$,
weak counterpart of (\ref{gm}), where $T^3_W$ is the weak isospin and $Y_W$ the 
weak hypercharge.
Like $Q$, $Y_W$ generates the $U(1)_Y$ symmetry group. Thus,
this yields the $SU(2)_L \times U(1)_Y$ gauge group.
The fundamental fermions, quarks and leptons, come in three generations or families with weak 
isospin and weak hypercharge quantum numbers as shown in the Table \ref{t3}.

\bt[b]
\caption{Assignment of 
quantum numbers in $SU(3)_L \times U(1)_Y$}
\begin{center}
\begin{tabular}{|c|c|c|c|c|} \hline
generation	& $T_L^3$ 	& $Y_L$ 	&  $T_R^3$ & $Y_R$\\ \hline
u $\hspace{0.3cm}$ 
c $\hspace{0.3cm}$ 
t 		& 1/2		& 1/6		& 0		& 2/3 \\
d $\hspace{0.3cm}$ 
s $\hspace{0.3cm}$ 
b 		& -1/2		& 1/6		& 0		& -1/3 \\
$\nu_e \hspace{0.3cm}$ 
$\nu_\mu \hspace{0.3cm}$ 
$\nu_\tau$ 	& 1/2		& -1/2		& 0		& 0 \\
$e \hspace{0.3cm}$ 
$\mu \hspace{0.3cm}$ 
$\tau$ 		& -1/2		& -1/2		& 0		& -1 \\ \hline
\end{tabular}
\label{t3}
\end{center}
\et

\subsubsection{Electroweak Lagrangian}
The electroweak Lagrangian of the non-Abelian local gauge
$SU(2)_L \times U(1)_Y$ symmetry is of the form
\bea
{\cal L}_{ew} &=& -{1 \over 4} W^{i,~\mu \nu}W^i_{\mu \nu} 
		-{1 \over 4} B^{\mu \nu}B_{\mu \nu}
		+2\sum_{f=1}^3 \bar \Psi^f_L\gamma_\mu D^\mu \Psi^f_L\\
D^\mu &=& \partial^\mu + i g t^i W^{i, \mu} + i g^\prime Y B^\mu  \label{lew}
\eea
where $g$ and $g^\prime$ are coupling constants. We get the photon field and the 
$Z^0$ field as 
the linear combination of $W^3$ and $B$ coupled to the electromagnetic current:
\bea
A_\mu &=& - \sin \theta_W W^3_\mu + \cos \theta_W B_\mu, \\
Z_\mu &=& \cos \theta_W W^3_\mu + \sin \theta_W B_\mu
\eea 
where 
$\tan \theta_W = {g^\prime / g}, \hspace{0.15cm} e = g \sin \theta_W$
The interaction part is
\be
{\cal L}_I = - \left\{{g \over {2\sqrt{2}}}J_\mu^+ W^{-,\mu}
			+{g \over {2\sqrt{2}}}J_\mu^- W^{+,\mu}\\
                      +eJ_\mu^{em}A^\mu \right\}
\ee
The currents are given by
\bea
J_\mu^+ &=& \sum \bar q \gamma_\mu (1-\gamma_5) V_{CKM} q
			+ \sum \bar l \gamma_\mu (1-\gamma_5) l
			\label{jw}\\ 
J_\mu^{NC} &=&  \sum \bar \Psi \gamma_\mu (v-a \gamma_5) \Psi\\
J_\mu^{em} &=&  \sum \bar \Psi \gamma_\mu Q \Psi
\eea
where $V_{CKM}$ is the CKM matrix of flavour mixing (see below) and 
$v = T^3_L - 2 Q \sin^2 \theta_w, a = T^3_L$.

\subsubsection{Spontaneous breaking of symmetry}
In the electroweak Lagrangian in (\ref{lew}), introducing a mass term for both fermions
and gauge bosons is forbidden by gauge invariance. 
However, the Higgs mechanism of spontaneous breaking
of the $SU(2)_L \times U(1)_Y$ symmetry \cite{hig} gives mass for the fermions and
the three bosons of weak interaction. For the Higgs doublet field, $\Phi$,
\be
\Phi =
\left(
\begin{array}{l}
\phi^+ \\
\phi^0
\end{array}
\right)
\ee
the Lagrangian is of the form
\bea
{\cal L}_\phi(x) &=& (D^\mu \Phi)^{\dag} (D_\mu \Phi) 
        - \sum_{\chi} f_\chi \bar \chi_L \tilde{\Phi} \chi_R 
        + \mu^2 \Phi^* \Phi - \lambda (\Phi^{\dag} \Phi)^2\label{lphi}\\
D_\mu \Phi &=& ({\bf I}(\partial_\mu + i {g \over 2}B_\mu) + 
		i g^\prime{\tau_i \over 2}W_\mu)\Phi
\eea 
In (\ref{lphi}), the three terms stand for free, Higgs-fermion interaction
and potential parts
as in the order given.
Minimising the potential, ground state Higgs field is 
\be
\Phi = 
\left(
\begin{array}{l}
0\\
v /\sqrt{2}
\end{array}
\right), \hspace{0.5cm} v = \sqrt{{\mu^2 \over \lambda}}
\ee
This non-vanishing field describes the spontaneous breaking
of the electroweak symmetry. Thus, the masses of fermions and bosons
are generated:
\bea
{\cal L}_{mass} = &-&{v \over {\sqrt{2}}}\sum f_a \bar a a
		  +\left({vg^\prime \over 2}\right)^2W_\mu^+ W^{-,\mu}\nonumber\\
		  &+&{v^2 \over 8}(W_\mu^3 \hspace{0.35cm} B_\mu)
		  \left(
		  \begin{array}{ll}
		  {g^\prime}^2 & -gg^\prime\\
		  -gg^\prime & g^2
		  \end{array}\right)
		  \left(
		  \begin{array}{l}
		  W^{3, \mu}\\
		  B^\mu
		  \end{array}\right)
\eea
where $a$ stands for all the fermions, excluding the neutrinos.

Masses for the bosons and the fermions (but not neutrinos) are, then, given by
\bea
m_a &=& {v \over \sqrt{2}}f_a, \hspace{1.0cm} a = u, d, e,...\\
M_W &=& {v \over 2}g^\prime, \hspace{1.2cm}
M_Z = {v \over 2}\sqrt{g^2+{g^\prime}^2}\\
M_\gamma &=& 0
\eea

\subsubsection{Quarks mixing}
Mixing of different flavours is caused by the charged current
mediated by $W^\pm$. The quark-mixing matrix, $V_{CKM}$, in (\ref{jw})
is a unitary (3 $\times$ 3) matrix \cite{ckm}.
\be
V_{CKM} =
\left(
\begin{array}{lll}
V_{ud} & V_{us} & V_{ub}\\
V_{cd} & V_{cs} & V_{cb}\\
V_{td} & V_{ts} & V_{tb}
\end{array}
\right)
\label{Vkm}
\ee
By the above equation, the weak basis, $q^\prime$ and mass basis $q$ 
are related by
$\bar q_i^\prime = V_{ij} q_j$
The unitarity of the matrix $V$ is: 
$V_{ij} \ra exp(i\phi_i)V_{ij}exp(-i\phi_j)$.

Further, the elements of the CKM matrix satisfy unitarity constraints: 
any pair of rows or columns are orthogonal. There
are six constraints (\ref{Vkm}). For example, 
\be
V_{ud}V_{td}^*+V_{us}V_{ts}^*+V_{ub}V_{tb}^* = 0
\ee
This is useful to determine one element from the others.

\subsection{Weak Decays of Heavy Hadrons}
Knowledge of the weak decays of beauty hadrons, especially $B$ mesons,
is an important requirement to test the Standard Model \cite{ali}. Of about
twenty input parameters, nine are the CKM matrix elements. The unitarity
test of the CKM matrix demands precise prediction of its elements.
The determination of $V_{ub}, V_{cb}, V_{tb}, V_{ts}$, 
and $V_{td}$, involves  the $b$ hadron weak processes: the charged current
mediated modes, {\it viz.,} the dominant $b \ra c + Y$, and the CKM suppressed,
$b \ra u + Y$ determine
$V_{cb}$ and $V_{ub}$ respectively; the rare decays of the neutral current,
$b \ra s + Y$ and $b \ra d + Y$ respectively determine $V_{ts}$ and $V_{td}$;
and the weak decay, mediated by the charged current, $t \ra bW$,
determines $V_{tb}$. ($Y$ denotes either a lepton pair or a quark pair
or a photon/gluon.). Precise determination of the elements of the CKM matrix
is stringent need for checking the CKM model of
the CP violation in the SM \cite{bigi}. 

The heavy hadrons, hadrons containing a heavy quark whose mass $m_Q \gg 
\Lambda_{QCD}$, is easier to study from the QCD point of view. 
For, the coupling constant is smaller
at $m_Q$ and the Compton wavelength of the heavy quark, $1/m_Q$, is much
smaller than $\Lambda_{QCD}^{-1}$. Thus, a lot of simplification arises
in describing the properties and decays of these hadrons. In the limit,
$m_Q \ra \infty$, with the resultant so called heavy quark symmetry, we 
get a low energy description of these heavy hadrons. This enables a
model independent prediction of the CKM matrix elements $|V_{ub}|$
and $|V_{cb}|$.
\newpage
\section{Hadrons with Spin-flavour Symmetry}
In this section, we discuss the spin-flavour symmetry or heavy quark symmetry
exhibited by the hadrons 
containing a heavy quark, quark of mass $m_Q \gg \Lambda_{QCD}$. Then, we describe the
theory of inclusive decays of heavy hadrons\cite{neub,neub0,shif,big}.
\subsection{Spin-flavour symmetry}
Because the heavy quark mass can be taken to the asymptotic limit, the study
of weak interaction of the heavy hadrons becomes simplified \cite{shur,nw,vol,iw}:
\begin{itemize}
\item
the asymptotic freedom: the coupling constant is small enough to
allow perturbative calculations;

\item
the Compton wavelength of the heavy quark, $1/m_Q$, is much
greater than the typical hadron size, $1/\Lambda_{QCD}$.

\end{itemize}

In the limit $m_Q \ra \infty$, the heavy quark acts as a static source of
colour inside the hadron. Since the light degrees of freedom cannot resolve
the distances smaller that the typical hadron size, the light stuff
interaction with the heavy quark vanishes. Consequently, 
either state of the heavy quark spin is independent of the other.
Thus heavy quark spin generates the $SU(2)$ spin symmetry.
On the other hand, the replacement of an infinitely heavy quark by another
infinitely heavy quark does not change the description of the system. This leads
to the flavour symmetry. Hence, for $N_f$ heavy flavours, the hadron concerned
is described by $SU(2N_f)$ symmetry, called spin-flavour or heavy quark
symmetry.

In the static limit of the heavy quark, the four-velocity of the heavy
quark is $v_\mu = (1, {\bf 0})$, {\it i. e.,} the heavy quark is in the rest
frame of the hadron containing it. Thus, both the decaying hadron and the
corresponding final state are looking the same in this limit, since
the product of the four-velocities of the initial and final state heavy 
quarks is unity.

Also the third components of the heavy quark spin, $\pm 1/2$ and that
of the light quark(s), $j$ are separately conserved, $J = j \pm 1/2$.
With the heavy quark symmetry, the heavy hadrons can be classified
by the quantum numbers of the light degrees of freedom, {\it viz.,}
flavour, spin, parity, etc., but independent of heavy quark spin and 
mass \cite{iw1}.
As will be described in more detail later, the mass of the hadron
is given by an expansion in the inverse powers of the heavy quark mass:
\be
M_H = m_Q + \bar \Lambda + O(1/m_Q) \label{be}
\ee
where $\bar \Lambda$ stands for the energy of the light degrees of freedom. 
The term $O(1/m_Q)$ is due to the kinetic energy of the heavy quark 
inside the hadron and the hyperfine mass splitting due to the
spin orientation of the heavy quark with the light degrees of freedom.
Its magnitude depends on the heavy quark mass, of the order of 
$\Lambda_{QCD}$.

Under the breakdown of the $SU(3)$ flavour symmetry, the $c$ and $b$ meson
mass differences are of approximately same magnitude:
\bea
m_{B_s} - m_{B_d} = \bar \Lambda_s - \bar \Lambda_d \\
m_{D_s} - m_{D_d} = \bar \Lambda_s - \bar \Lambda_d 
\eea
where the mass differences are respectively given by $90 \pm 3$
and $99 \pm 1$ MeV. 
Similarly, the mass difference between the baryon and the meson
is given by:
\bea
m_{\Lambda_b} - m_B = \bar \Lambda_{\Lambda_b} - \bar \Lambda_{B} \\
m_{\Lambda_c} - m_D = \bar \Lambda_{\Lambda_c} - \bar \Lambda_{D}
\eea
Their values are predicted to be $346 \pm 6$  and $416 \pm 1$ MeV respectively.
These mass differences receive corrections of $O(\lq/m_Q)$.

\subsection{Effective Theory}
In the limit, $m_Q \ra \infty$, the QCD Lagrangian turns out to be
a low energy effective one as the heavy particle is irrelevant at low
energies \cite{ef,eh,gr,hg,fg,fgl,am}. 
Now the theory is similar to the Fermi theory of weak interaction 
discussed in the previous section.

As mentioned earlier, in the heavy quark limit, the heavy quark is in the rest 
frame of the parent hadron and it is almost on-shell. Its momentum is
\be
p_Q^\mu = m_Q v^\mu +k^\mu 
\ee
where $k$ is the residual momentum of $O(\lq)$. Accordingly, we define the
propagator and the vertex function as
\be
{i \over {p_Q^\mu - m_Q}} = {i \over {v.k}}{1+ {\not\!v} \over 2}
\hspace{0.3cm}~\mbox{and} \hspace{0.3cm} ig_s v^\alpha T_a
\label{feyn}
\ee
respectively. In the above equation, $(1\pm {\not\!v})/2$ are the positive and negative 
projection operators. They act finitely on the large and small components 
of the fields respectively:
\be
h_v(x) = e^{im_Q v.x} P_+ Q(x),
\hspace{0.9cm} H_v(x) = e^{-im_Q v.x} P_- Q(x)
\ee
such that 
\be 
Q(x) = e^{-i m_Q v.x}[h_v(x)+H_v(x)]
\ee
Because of the projection operator, the field defined above satisfies
${\not\!}vh_v = h_v$ and ${\not!v}H_v = -H_v$, where $h_v$ annihilates 
a quark of velocity 
$v$ and $H_v$ creates an heavy antiquark. 

With these new fields, the QCD Lagrangian
\be
{\cal L}_{QCD} = \bar Q(x) (\not\!D - m_Q) Q(x)
\ee
turns out to be of the following form for a heavy quark
\be
{\cal L}_Q = \bar h_v iv.D h_v - \bar H_v (iv.D + 2m_Q) H_v
+\bar h_v (i \not\!D_\perp) H_v +\bar H (i \not\!D_\perp) h_v
\ee
where $D^\mu_\perp = D^\mu - v^\mu$, $v.D$ satisfying $v.D_\perp$ = 0 in the rest frame,
the $h_v$ term corresponds to the massless degrees of freedom and the $H_v$ term to
the heavy degrees of 
freedom which have to be integrated out, with fluctuation being twice the heavy quark mass. 
By using the equation of motion, the heavy degrees of freedom, $H_v$,
can be eliminated. That is by taking derivative of the above equation with
respect to $\bar H_v$, we get the differential equation and its solution
as
\be
(iv.D + 2m_Q)H_v = i \not\!D_\perp h_v, \hspace{0.5cm} H_v = 
{1 \over {2m_Q + iv.D}}i \not\!D_\perp h_v
\ee
Thus the nonlocal effective action results in 
\be
{\cal L}_{eff} = \bar h_v iv.D h_v + \bar h_v i \not\!D_\perp {1 \over {2m_Q+iv.D}} 
i\not\!D_\perp h_v
\ee
where the second term corresponds to virtual processes. This Lagrangian can be 
expanded in powers of $1/m_Q$:
\be
{\cal L}_{eff} = \bar h_v iv.D h_v + 
{1 \over {2m_Q}} \sum_{n=0}^\infty \bar h_v i\not\!D_\perp \lln {-iv.D \over {2m_Q}} 
\rln i\not\!D_\perp h_v
\ee
The Lagrangian is explicitly given as
\be
{\cal L}_{eff} = \bar h_v iv.D h_v + 
{1 \over {2m_Q}} \bar h_v (iD_\perp)^2 h_v + {g_s \over {4m_Q}} 
\bar h_v \sigma_{\mu \nu} G^{\mu \nu} h_v + O(1/m_Q^2)
\ee
Thus, in the limit $m_Q \ra \infty$, only the first term survives, which is
independendent of the heavy quark mass:
\be
{\cal L}_{HQET} = \bar h_v iv.D h_v  \label{heff}
\ee 
For this Lagrangian, the Feynman rules are given already in (\ref{feyn}).
In the above, we have applied the heavy quark limit to the effective
Lagrangian, not the QCD Lagrangian. Thus, the heavy quark symmetry is
not the symmetry of QCD but that of the low energy effective theory, 
(\ref{heff}). The corrections appear to ${\cal L}_{HQET}$
due to the finite mass of the heavy quark. As the heavy quark mass increases,
the expansion becomes more and more powerful.

In the rest frame, the $O(1/m_Q)$ operators are kinetic and chromomagnetic 
operators.
\be
O_{kin} = {1 \over {2m_Q}} \bar h_v (iD_\perp)^2 h_v \ra 
-{1 \over {2m_Q}}\bar h_v (i \vec{D})^2 h_v
\ee
which is the gauge covariant extension of the kinetic energy due to heavy quark motion.
\be
O_{mag} = {g_s \over {4m_Q}} \bar h_v \sigma_{\mu \nu} G^{\mu \nu} h_v 
 \ra -{g_s \over {m_Q}}\bar h_b \vec{S} \vec{B_c} h_v
\ee
is the non-Abelian analogue of the Pauli interaction describing the colour magnetic
coupling of the heavy quark spin to the gluon field. In the
above equation
for the magnetic operator, the spin operator $\vec{S}$ and the component of the
colour magnetic field $B^i_c$ are given by
\bea
S^i &=& {1 \over 2} \left(
\begin{array}{ll}
\sigma^i & 0 \\
0 & \sigma^i
\end{array}
\right), \hspace{0.5cm}
[S^i, S^j] = i \epsilon^{ijk}S^k\\
B_c^i &=& -{1 \over 2} \epsilon^{ijk}G^{jk}
\eea

The corrections appearing at the next-to-leading order and beyond can be expressed
as the short and long distance parts entangled. While the short distance physics
is contained in the Wilson coefficients, the long distance physics are given by
the local operators.

As we have mentioned earlier (\ref{be}),  
the energy of the light degrees of freedom
$\bl$ varies with the heavy quark mass. But the choice of the heavy quark mass
is not unique. The mass previously referred to in this section is the 
mass in the Lagrangian.  If one chooses it to be the pole mass, $m_Q$,
then $\bl$ is given in the asymptotic limit of the heavy quark mass
as
\be 
\bl = (M_H - m_Q)|_{m_Q \ra \infty}
\ee
If we define the heavy quark mass in some other scheme and at a certain scale,
then $\bl$ changes accordingly. However, if defined alternately as
\be
\dl = (M_H - \bar m_Q)
\ee
then the quantity $\dl$ is invariant, where $\delta m$ is known as the residual 
mass of the order of $\lq$ \cite{fn0}. In that case, since the HQET allows the residual 
mass in the leading Lagrangian, ${\cal L}_{HQET}$ would be of the form
\be
{\cal L}_{HQET} = \bar h_v iv.D h_v - \delta m \bar h_v h_v
\ee
However, if the heavy quark mass is defined as the pole mass, then
the residual mass term, $\delta m = 0$. Now, these three quantities
$m_Q, \bl$ and $\delta m$ are non-perturbative parameters of HQET.
They cannot be defined precisely in perturbation theory where ambiguities
are  of
the order of $\lq$ given by the renormalon graphs.

\subsection{Exclusive decays}
The heavy-quark symmetry nicely describes the exclusive weak decays
by a single function \cite{nw,vol,iw}. With a mass-independent normalisation of
the meson states,
\be
\ln M(p^\prime)|M(p) \rn = 2 (p^0/M_M)(2 \pi)^3 \delta^3(p-p^\prime)
\ee
Thus, the meson state is redefined as 
\be
|M(v) > = M_M^{-1}| {\tilde M}(p) >.
\ee
Now, the meson state is characterised by the configuration of the
light degrees of freedom and this is the eigenstate of the effective 
Lagrangian which is supplemented by the QCD Lagrangian for the light
degrees of freedom.

Thus weak decays of a heavy meson state, in the heavy quark limit, corresponds
to the replacement of  the heavy quark, $Q(v)$, in the initial state by another
$Q^\prime (v^\prime)$. The process is described by
\be
\ln M(v^\prime)|\bar h_v \gamma^\mu h_v | P(v) \rn = \xi(v.v^\prime)
(v+v^\prime)^\mu
\ee
where the form factor $\xi(v.v^\prime)$, is independent of the 
heavy quark mass.

Now, we see the weak decays of the heavy hadrons.
There are six form factors associated with transitions of one
flavour to another. The heavy quark symmetry relates them to one unique and 
universal function, known as Isgur-Wise function \cite{iw}, $\xi(v.v^\prime =1)$, 
where $v, v^\prime$ are the initial and final four-velocities. For
the transitions due to the currents, $V_\mu, A_\mu$, we have
\bea
\ln D^*(v^\prime) | A_\mu | B (v) \rn &=& h_{A_1}(y)(y+1)\epsilon^*\nonumber\\
&&-[h_{A_2}v^\mu+h_{A_3}(y)v^{\prime \mu}]\epsilon^*.v   \\
\ln D (v^\prime) | V_\mu | B (v) \rn &=&  h_+(y)(v+v^\prime)^\mu+
h_-(y)(v-v^\prime )^\mu \\
\ln D^* (v^\prime) | V_\mu | B (v) \rn &=& ih_v(y)\epsilon^{\mu \nu \alpha \beta}
\epsilon_\nu^* v_\alpha^\prime v_\beta  
\eea
The six form factors, in the heavy quark limit, are related to one another
by $\xi(y = v.v^\prime)$:
\bea
h_+(y) &=& h_v(y) = h_{A_1}(w) = h_{A_3} (w) = \xi(y),\nonumber\\
h_-(y) &=& h_{A_3}(y) = 0
\eea

Now consider the weak decay of $B$ into $D$ which amounts to the replacement
of the $b$ quark in the initial hadron by the $c$ quark. With the application 
of heavy quark symmetry, we have
\be 
{1 \over {\sqrt{M_B M_D}}}\ln D(v^\prime)|\bar c_{v^\prime} \gamma^\mu b_v
|B(v) \rn = \xi (y) (v+v^\prime)^\mu
\ee
The determination of the matrix elements by $\xi(y)$ is facilitated by  
comparison with 
\be
\ln D(v^\prime)|\bar c_{v^\prime} \gamma^\mu b_v
|B(v) \rn = f_+(q^2)(p+p^\prime)^\mu - f_-(p-p^\prime)^\mu.
\ee
Thus the independent form factors $f_\pm$ are given by $\xi(y)$.
\be
f_\pm (q^2) = {M_B \pm M_D \over {2 \sqrt{M_B M_D}}}\xi(y),
\hspace{0.2cm} q^2 = M_B^2 +M_D^2 - 2yM_B M_D.
\ee
If $\xi(y) = 1$, the form factors are given by the hadron mass. Thus,
the differential decay rate of $B \ra D l \nu$ is given by \cite{n3}
\be
{d \gm(\bar B \ra D l \nu) \over {dy}}
= {G_f^2 \over {48 \pi^3}}|V_{cb}|^2 (M_B+M_D)^2 M_D^5 (y^2 -1)^{3/2} \xi^2(y)
\label{exd}
\ee
The symmetry braking corrections, arising in the preasymptotic limit of
the heavy quark mass, are $O(\lq/M_Q)^2$. However, the matrix elements describing the $1/m_Q$ 
corrections vanish at the equal velocity limit. This is known as Luke's theorem \cite{lu},
an analogue of the Ademello-Gatto theorem \cite{ag}. The corrections of $O(1/m_Q^2)$ are studied
in \cite{fn,n3,n4}

Application of the formalism in which the weak exclusive decays are expressed
in terms of the Isgur-Wise function is to extract the elements of the CKM 
matrix. From (\ref{exd}), we can determine $|V_{cb}|$ in a model independent
way \cite{n5,jm,bg0,fg1,n6,n7,n8,ig2,n9,nr,n10,n11,bgl,bl}. 

\subsection{Inclusive decays}
The theory of the inclusive decays of the heavy hadrons is formulated using the
optical theorem \cite{cg,buv,bsuv,bk,mw,l2,tm,fln,n12,mn}. It
allows one to determine the quantities like the CKM matrix elements cleanly
in a model independent way. In this mode, considering the final state as
the sum of all the hadronic states, we eliminate the bound state effects which
are not known from the first principles of QCD. This is based on the idea of 
quark-hadron duality: the hadronic quantities can be calculated in terms of quarks and gluons
at a particular kinematic point. That is, the hadronic quantities specified in 
the physical region are calculable by means of a smearing mechanism in terms
of quarks/gluons \cite{pqw}. 
 
\begin{figure}
\begin{center}
 \epsfig{figure=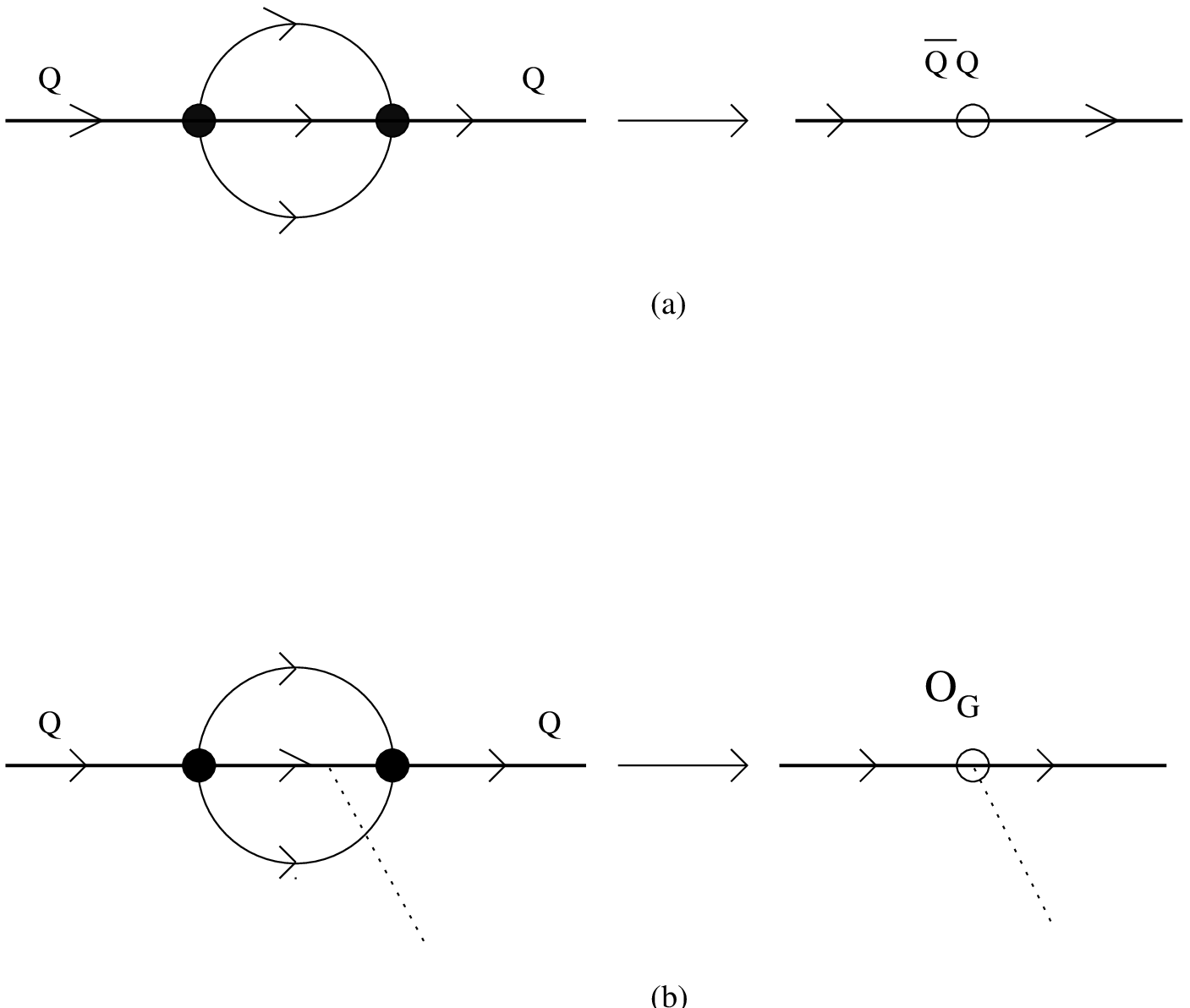}
 \caption{Perturbative corrections to the Transition operator, T, (left)
 and to the local operators in OPE (right). (a) $\bar Q Q$ (b)
 $\bar Q g \sigma G Q $ (Dotted lines are gluons).}
  \label{fig:1.1}
\end{center}
\end{figure}

Inclusive decay rate is expressed using the optical theorem as
\be 
\Gamma(H_b \ra X) = {1 \over M_{H_b}} Im \ln H_b|T|H_b \rn
\ee
where T is the transition operator given by
\be
T = i \int d^4x T\{{\cal L}_{eff}(x), {\cal L}_{eff}(0)\} \label{top}
\ee
The effective Lagrangian ${\cal L}_{eff}$ is given by
\be
{\cal L}_{eff} = -2 \sqrt{2} G_f J^\mu_{CC}J_{CC,~\mu}     \label{effl}
\ee
Insertion of a complete set of states between the time ordered product, 
the decay rate is of the form
\be
\Gamma(H_b \ra X) = {1 \over {2M_{H}}} \sum_X (2\pi)^4 \delta^4(p_H-p_X)
|\ln H_b | {\cal L}_{eff}|H_b \rn |^2
\ee
The Lagrangian (\ref{effl}) receives a short distance correction due to the virtualities
between $M_W$ and $m_b$. For the transition operator in (\ref{top}), the OPE can be 
constructed in terms of a series of local operators containing heavy quark fields.
These operators are arranged according to their dimensions.
\begin{itemize}
\item
The first operator is of dimensions three (D = 3): $\bar Q Q$ (see Figure
\ref{fig:1.1}a).

\item
Secondly, D = 4  operator $\bar Q iD Q$ is reduced to $m_Q \bar Q Q$ by
the equation of motion (see Figure\ref{fig:1.1}b). 

\item
Then the operator of D = 5 which is distinct from the D = 3 operator is the one involving
the gluon field: $\bar Q g \sigma_{\mu \nu}G^{\mu \nu} Q$
\end{itemize}

Then the inclusive decay rate is given by these set of operators by
\be
\Gamma(H_Q \ra X) = {G_f^2 |V_{KM}|^2 m_Q^5 \over {192 \pi^3}}
\lln c_3 \ln \bar Q Q \rn_H + c_5 {\ln \bar Q g \sigma_{\mu \nu}G^{\mu \nu} Q \rn_H 
\over {m_Q^2}} + \cdots \rln
\label{dr}
\ee
where $c_n$ are the short distance Wilson coefficients calculable in perturbation theory,
and
\be
\ln O \rn_H = { 1 \over {2M_H}} \ln H_Q | O |H_q \rn
\ee
The hadronic matrix elements are expanded in inverse powers of the heavy 
quark mass (Figure 1.1) as
\bea
\ln \bar Q Q \rn_H &=& 1 - {\mu_\pi^2(H) - \mu_G^2(H) \over {2m_Q^2}}+O(1/m_Q^3)\\
\ln \bar Q g \sigma_{\mu \nu} G^{\mu \nu} Q \rn_H &=& 2\mu_G^2(H)+O(1/m_Q^2)
\eea
where the HQET matrix elements are defined as
\bea
\mu_\pi^2(H) &=& {1 \over {2M}}\ln H(v)|\bar Q (iD)^2 Q|H(v)\rn \\
\mu_G^2(H) &=& {1 \over {2M}}\ln H(v)|\bar Q {g \over 2} \sigma_{\mu \nu} G^{\mu \nu} Q|H(v)\rn
\eea
Hence, we have the decay rate of the form
\be
\Gamma(H_Q \ra X) = {G_f^2 |V_{KM}|^2 m_Q^5 \over {192 \pi^3}}
\left\{ c_3 \lln 1 - {\mu_\pi^2(H) - \mu_G^2(H) \over {2m_Q^2}} \rln +
2c_5 {\mu_G^2(H) \over {m_Q^2}}+ \cdots \right\}
\label{drf}
\ee

In the above equation, the leading order term corresponds to the free quark decay rate. At $O(1/m_Q^2)$,
the term $\mu_\pi^2(H)$ describes the motion of the heavy quark inside the hadron and the term
$\mu_G^2(H)$ the hyperfine interaction. The kinetic term for a $B$ meson is estimated to be in the range 
of 0.1 to 0.5 GeV$^2$ \cite{n2,bs1,fl,bb00} and the latest value is given by
\be
\mu_\pi^2(B) = -\lambda_1 = 0.3 \pm 0.2  ~\mbox{GeV$^2$}
\ee
For the $\Lambda_b$ baryon, it is obtained using the mass values of the hadrons from the
following relation
\be
(m_{\Lambda_b} - m_{\Lambda_c}) - (\overline m_B - \overline m_D) =
[\mu_\pi^2(B) - \mu_\pi^2(\Lambda_b)]\lln {1 \over {2m_c}} - {1 \over {2m_b}}\rln + O(1/m_Q^2)
\ee
where $\overline m_{B,D}$ are the spin-averaged masses of $B$ and $D$ mesons. The kinetic energy 
of a heavy quark inside a meson and a baryon differs by 
\be
\mu_\pi^2(B) - \mu_\pi^2(\Lambda_b) = 0.01 \pm 0.03 ~\mbox{GeV$^2$}
\ee
It should be noted that the bound on the two terms are model independent
\be
\mu_\pi^2(B) > \mu_G^2(B) \approx 0.35 ~\mbox{GeV$^2$}
\ee
 
On the other hand, the hyperfine interaction term is determined precisely from the difference
between the masses of the vector and pseudoscalar mesons:
\be
\mu_G^2(B) = {3 \over 4}(M_{B*} - M_B) = 0.12 ~\mbox{GeV$^2$}
\ee
For baryons it vanishes except in the case of the $\Omega_Q$ baryon.

The phase space factors corresponding to the $C_n$ are given by
\bea
c_3 &=& 1-8x+8x^3-x^4-12x^2 \log x^2 \label{c3}\\
c_5 &=& (1-x)^4 \label{c5}
\eea
where $x = m_q^2/m_b^2$, the mass ratio of the final state quark to the decaying quark.  
The leading order factor given in (\ref{c3}) is exact \cite{nir} and the other 
due to higher orders in (\ref{c5}) is known only partially \cite{lsw}. 

The total decay rate scales like $m_b^5$. The leading order term corresponds to
the free heavy quark decay rate. This is universal for all hadrons of a given flavour.
The next-to-leading order terms appear at $O(1/m_b^2)$. They describe the heavy quark
motion inside the hadron and the hyperfine interaction due to heavy quark spin projection.
However, the next to leading order contributions are suppressed by $m_b^{-2}$. 

With these, the important predictions for inclusive properties of beauty hadrons
are given in Table \ref{t31}.

\bt
\begin{center}
\caption{HQE predictions vs Experimental Values}
\begin{tabular}{|c|c|c|} \hline
				& theory 		& experiment\\ \hline
$\tb$$(B^+)$/$\tb$$(B^0_d)	$	& 1+$O(1/m_b^3)$ 		&1.02 	$\pm $0.04 \\
$\tb$$(B^0_s)$/$\tb$$(B^0_d)	$	& (1.0 $\pm$ 0.01)+$O(1/m_b^3)$ &1.01 	$\pm $0.07 \\
$\tb$$(\Lambda_b)$/$\tb$$(B^0_0)$	& 0.98 + $O(1/m_b^3)$		&0.78 	$\pm $0.05 \\
$Br(B)_{SL}$ 			& 12.5\%				& 10.5\% \\
$n_c$ 				& 1.16 				& 1.24 \\ \hline
\end{tabular}
\label{t31}
\end{center}
\et

Because of the discrepancy in the case of $\Lambda_b$, the semileptonic branching ratio of $B$ mesons,
$Br(B)_{SL}$ \cite{bbsv,bbbg,bbfg} and the charm counting, $n_c$, \cite{ap,bdy},
the third order in $1/m_Q$ terms have necessarily
to be evaluated. This will be discussed in the next chapter.

\chapter{Four-quark operators}
In this Chapter we discuss the evaluation of the expectation
values of four-quark operators (EVFQO). The EVFQO are estimated in two different approaches:
(1) in terms of the form factor characterising the light scattering off the heavy quark
inside the hadron, using a potential model\footnote{S. Arunagiri, {\it Colour-straight four-quark 
operators and lifetimes of $b$-flavoured hadrons}
\cite{arun1}.},
and (2) from the difference in the total decay rates
of beauty hadrons\footnote{S. Arunagiri, {\it A note on spectator effects and quark-hadron duality in 
inclusive beauty decays} \cite{arun2}.}. 

\section{Preview}
As already mentioned, in order to systematically
accommodate the contribution coming from the terms at the
third order of the heavy quark expansion, it is necessary
to evaluate the expectation values of the four-quark operators.
These are the dimension six operators involving both the heavy and
light quark fields. Though their contributions are negligible as
compared to the volume of heavy quark occupation,
$O{(\Lambda_{QCD}/m_Q)^3}$, they are considerably enhanced due to
partial compensation by the four-quark
phase-space. As previously noted, the lifetimes of all mesons
as well as baryons are almost equal among themselves respectively. But
the knowledge of their individual hierarchy depends on the outcome
of the terms at $O(1/m_Q^3)$. Besides, the persisting confrontation of
the theoretical estimate of the lifetime of $\Lambda_b$ with
the experimental
value necessiates the evaluation of the four-quark operators.

The hadronic matrix elements at $O(1/m_Q^3)$ is
\be
C(\mu)\left<H_b|(\bar Q \Gamma_\mu Q)(\bar q \Gamma^\mu q)|H_b\right> \label{m3}
\ee
where $\Gamma_\mu$ stands for some arbitrary Dirac structure,
the coefficient $C(\mu)$ describes the short distance
QCD processes which will be discussed in the next chapter. 
The operators, in (\ref{m3}), are induced into a set of four by
the spectator effects. They are:
\bea
O^q_{V-A} &=& \bar b \gamma_\mu(1-\gamma_5) q \bar q \gamma^\mu(1-\gamma_5)b\\
O^q_{S-P} &=& \bar b (1-\gamma_5) q \bar q (1+\gamma_5)b\\
T^q_{V-A} &=& \bar b \gamma_\mu(1-\gamma_5)t^a q \bar q \gamma^\mu(1-\gamma_5)t^ab\\
T^q_{S-P} &=& \bar b (1-\gamma_5)t^a q \bar q (1+\gamma_5)t^ab
\eea
By insertion of vacuum states, we have 
\bea
\left<O^q_{V-A}\right>_{B_q} &=& \left<O^q_{S-P}\right>_{B_q} = f_B^2 m_B^2\\
\left<T^q_{V-A}\right>_{B_q} &=& \left<T^q_{S-P}\right>_{B_q} = 0
\eea
where $f_B$ is the leptonic decay constant of $B$ meson. The evaluation of
the operators is a difficult task. It involves the nonperturbative aspect of
QCD. The expectation values of these operators are understood
as the probability of finding both the quarks at the origin,
$|\Psi(0)|^2$.
The traditional way of estimating the expectation values of these
operators for a meson is to use the vacuum insertion approximation and
express
the value in terms of the leptonic decay constant:
\be
{1 \over {2m_B}}
\left<B|(\bar Q \Gamma_\mu Q)(\bar q \Gamma^\mu q)|B\right>
=|\Psi(0)|^2_B = {f_B^2 m_B \over {12}}
\ee
On the other hand, for baryon, one has to use the valence quark
approximation.
These estimations are model dependent. We next discuss the
evaluation of the expectation values of four-quark operators
made by Rosner \cite{rosner96}, Neubert and Sachrajda \cite{neub97}
and De Fazio and Colangelo\cite{colan96},
followed by the evaluation made in potential model and from the
difference in the decay rates of the $SU(3)$ triplet mesons and baryons.

In \cite{rosner96}, Rosner estimated the wavefunction density for $\lbb$
in constituent model using the DELPHI value of mass difference between
$\Sigma$ and $\Sigma^*$ \cite{abreu95}.
With the four-quark operators of a meson estimated in terms of the leptonic
decay constant, the operators of a baryon can be made in a quark model
using the hyperfine mass splitting of the $b$-baryons.
The mass splittings due to the hyperfine interaction in a meson, $M_{i \bar
j}$,
and in a baryon, $B_{ijk}$, are given by
\bea
\Delta M(M_{i \bar j}) &=& { 32 \pi \over 9} \alpha_s {\left< {\hat
s}_i.{\hat s}_j
\right> \over {m_i m_j}}|\Psi(0)|^2_{i \bar j}\\
\Delta M(B_{ijk}) &=& { 16 \pi \over 9} \alpha_s
\sum_{i>j} {\left< {\hat s}_i.{\hat s}_j
\right> \over {m_i m_j}}|\Psi(0)|^2_{ij}
\eea
The value of $\left< {\hat s}_Q.{\hat s}_{\bar q} \right>$ is (1/4, -3/4)
for a $(^3 S_1, ^1S_0)$ $Q \bar q$ meson. With $S_{qq} = 1$,
$\left< {\hat s}_Q.{\hat s}_{\bar q} \right>$ is (1/4, -1/2)
for states with total spin $(3/2, 1/2)$ for a $Qqq$ baryon. With these,
the expectation values of four-quark operators for a $b$-baryon is:
\be
|\Psi(0)|^2_{\Lambda_b} ={4 \over 3}  {\Delta B_{ijk} \over
{\Delta M_{ij}}}
|\Psi(0)|^2_B
\ee

For the baryon mass splitting, $M(\Sigma_b^*)-M(\Sigma_b) = 56 \pm 16$ MeV \cite{abreu95}
and for $M(B^*)-M(B) = 46$ MeV, we get
\be
|\Psi(0)|^2_{\Lambda_b} = (2.6 \pm 1.3) \times 10^{-2}~~\mbox{GeV$^3$}
\ee
As will be discussed in the next chapter, this value is found
not sufficient to account for the smaller lifetime of $\Lambda_b$.
However, this work is in favour of larger value of $|\Psi(0)|^2$ for
$\lbb$ than that of $B$.

Neubert and Sachrajda \cite{neub97} parameterised the dimension six matrix elements
in terms of hadronic parameters.
The four-quark matrix elements between hadrons are
given in terms of the hadronic parameters $B_i, \epsilon_i$. For $B$
mesons
\bea
{1 \over {2m_B}}\ln B| (\bar b_L \gamma_\mu q_L)(\bar q_L \gamma^\mu b_L)
|B \rn \equiv {f_B^2 m_B \over 8}B_1\\
{1 \over {2m_B}}\ln B| (\bar b_R q_L)(\bar q_L b_R)|B \rn
\equiv {f_B^2 m_B \over 8}B_2\\
{1 \over {2m_B}}\ln B| (\bar b_L \gamma_\mu t^a q_L)
(\bar q_L \gamma^\mu t^a b_L)|B \rn
\equiv {f_B^2 m_B \over 8}\epsilon_1\\
{1 \over {2m_B}}\ln B| (\bar b_R t^a q_L)(\bar q_L t^a b_R)
|B \rn \equiv {f_B^2 m_B \over 8}\epsilon_2
\eea
where the hadronic parameters in the large-$N_c$ limit
\be
B_i = O(1), \hspace{0.5cm} \epsilon_i = O(1/N_c)
\ee
Using QCD sum rules, it is found that $\epsilon_1 = -0.15$ and
$\epsilon_2 = 0$.

Similarly, the matrix elements between $\Lambda_b$ states are expressed as
\be
{1 \over {2m_{\Lambda_b}}}\ln \lbb|
(\bar b_L \gamma_\mu q_L)(\bar q_L \gamma^\mu b_L)
|\lbb \rn \equiv -{f_B^2 m_B \over 8}r
\ee
where $r = {|\Psi(0)|^2_{\lbb}/{|\Psi(0)|^2_B}}$.
It is expected that the ratio is about unity. But Rosner argued that
it should be larger than unity.

\section{Potential Model Evaluation}

In the nonrelativistic quark theory, the wave function density
and diquark density are related to the associated operator 
\begin{equation}
O_6 = (\bar b_i \Gamma_b b^i)(\bar q_j \Gamma_q q^j), \label{op6}
\end{equation}
where the ${\Gamma_{b, q}}$ are arbitrary Dirac structures,
through
\bea
{1 \over {2M_B}} <B| (\bar b b)(\bar q q)| B> &=& |\Psi (0)|^2\\
{1 \over {2M_{\Lambda_b}}} <{\Lambda_b}| (\bar b b)(\bar q q)
|{\Lambda_b}>
&=& \int d^3y |\Psi (0, y)|^2
\eea
for a meson and a baryon respectively. The operators in (\ref{op6})
are colour singlet. The expectation values of
these operators are related to the
transition amplitude of the light quark scattering off the
heavy quark. Thus the determination is based on the
knowledge of the light quark scattering form factor. 

The wave function at the origin, in momentum representation,
is given by
\begin{equation}
\Psi(0) = \int {d^3{\bf p} \over {(2 \pi^3)^3}} \Psi({\bf p}).
\end{equation}
The transition amplitude is then the Fourier transform of the
light quark density distribution:
\begin{equation}
F({\bf q}) = {1 \over {2M_H}}<H_b({\bf q})|\bar q q(0) |H_b(0)> = 
\int d^3{\bf x} \Psi^*({\bf x}) \Psi({\bf x}) e^{i{\bf q}{\bf x}}.
\end{equation}
Integrating over all {\bf q} yields the expectation value:
\begin{equation}
\int {d^3{\bf q} \over {(2 \pi)^3}} F({\bf q}) = |\Psi(0)|^2 = 
{1 \over {2M_H}}<H_b|\bar b b \bar q q(0) |H_b> 
\end{equation}
For any Dirac structure ${\Gamma}$, the light quark current
density and the light quark transition amplitude are given by:
\bea
J_\Gamma({\bf x}) &=& \bar q \Gamma q({\bf x})\\
A_\Gamma({\bf q}) &=& 
{1 \over {2M_H}}<H_b({\bf q})|J_\Gamma(0) |H_b(0)> 
\eea
where the J${_\Gamma}$(0) is gauge invariant operator, and it is not
required to be a bilinear. Thus, for spin-singlet operators,
we have
\be
<H({\bf q})|\bar b b  J_\Gamma (0) |H_b(0)> = 
\int {d^3{\bf q} \over {(2 \pi)^3}} <H_b({\bf q})|J_\Gamma(0) |H_b(0)> 
\label{m1}
\ee
And, similarly for spin-triplet operators, 
\be
<\ H({\bf q})|\bar b {\bf \sigma_k} b \bar J_\Gamma (0) |H_b(0)> = 
\int {d^3{\bf q} \over {(2 \pi)^3}} <S_k \ H_b({\bf q})
|J_\Gamma(0) |H_b(0)> \label{m2}
\ee
with S/2 being the b-quark spin operator and 
\begin{equation}
|S_k \ H_b> = \int d^3{\bf x} \bar b \sigma_k b({\bf x}) |\ H_b>
\end{equation}
Equations (\ref{m1}) and (\ref{m2}) represent the general structure of four-quark 
operators. The above operators are local  as required for by
the HQE in the sense that the light quark operators enter
at the same point as the heavy b-quark operators. These relations
hold equally for different initial and final
hadrons having different momenta smaller than m${_b}$.

Equation (\ref{op6})  resolves into spin-singlet and spin-triplet operators
for ${\Gamma_b}$ = 1 and ${\Gamma_b}$ = ${{\bf \gamma} \gamma_5
(= {\bf\sigma})}$ respectively. The light quark elastic scattering
is described by the form factor F(q$^{2}$). In Ref.
\cite{pirjol98}, the exponential ansatz and the two pole
ansatz are used for the form factor. Both of them lead to a
result which is the same for meson and  baryon.

\subsubsection{Model for form factor}
As will be discussed in the following sections, the expectation
values of the colour-straight operators are parameterised in
terms of a single form factor  for both the B-meson and ${\Lambda_b}$
baryon \cite{arun1}. The extraction of the form factor involves an 
assumption of
a function such that it satisfies the constraints on the form factor
that F(q${^2}$ = 0) is equal to the corresponding charge of
the hadron. Then the form factor has to be extrapolated into
the region where q${^2}$ ${>}$ 0. We take the hadronic wave
function of ISGW harmonic oscillator model \cite{grin89} for
the form factor. The wave functions of ISGW model are the eigenfunctions of
the orbital angular momentum L = 0 satisfying the overlapping
integral which is chosen as the form factor.
\begin{equation}
F({\bf q}^2) = N^2 exp[-q^2/2(\beta_f^2+\beta_i^2)] 
\end{equation}
where N is the normalisation constant given by 
${[2 \beta_f \beta_i/(\beta_f^2+\beta_i^2)]^{3/2}}$ and
${\beta}$'s are oscillator strengths. For the same initial
and final hadrons ($\beta_i = \beta_f = \beta$), the transition amplitude is \cite{arun1}
\begin{equation}
\int {d^3{\bf q} \over {(2 \pi)^3}} F(q^2) =
{\beta^3 \over {4\pi^{3/2}}} =
|\Psi(0)|^2
\end{equation}
From the above equation, which is the master equation for evaluation of the
EVFQO, it is obvious that the transition amplitude and
hence the expectation values of four fermion operators are
proportional to the third power of the oscillator strength
of the hadron.

The calculation of ${\beta}$'s can be made using a QCD inspired 
potential. In the present calculation, we use the potential for
a B-meson containing the Coulomb, confining and constant terms: 
\begin{equation}
V(r) = {a \over {r}} + br + c
\end{equation}
where a = -- 0.508, b = 0.182 GeV${^2}$ and c = -- 0.764 GeV. With
quark masses m${_q}$ = 0.3 GeV (treating m${_u}$ =
m${_d}$), m$_s$ = 0.5 GeV and m${_b}$ = 4.8 GeV, using a
variational procedure with the wave function given
in position space as, 
\begin{equation}
\Psi(0)_{1s} = {\beta^{3/2} \over \pi^{3/4}} e^{-\beta^2r^2/2} \label{wf}
\end{equation}
the oscillator strengths of the beauty mesons are obtained. The values are \cite{grin89} 
\bea
{\beta}_{B_q} = 0.4 ~\mbox{GeV}\\
{\beta}_{B_s} = 0.44 ~\mbox{GeV}
\eea

For the ${\Lambda_b}$ baryon, we use the same procedure as before
but a potential of the form
\begin{equation}
V(r) = {1 \over {2}} \left ({a \over {r}} + br +
\zeta^\prime r^2 + c \right )
\end{equation}
where $\zeta^\prime = \zeta \beta^2$, with $\zeta = 1 $ GeV and 
the r${^2}$ term is a harmonic oscillator term justifying
the consideration that the ${\Lambda_b}$ be a two body system using
the same wave function as in (\ref{wf}), we determine the 
values of $\beta$ for the baryons \cite{arun1}:
\bea
{\beta_{\Lambda_b}} = 0.72 ~\mbox{GeV} \\
{\beta_{\Xi_b}} = 0.76 ~\mbox{GeV} 
\eea
We have taken the mass of the diquark system as 0.6 and 0.8 GeV for 
$\Lambda_b$ and $\Xi_b$ respectively.

The large value for the $\beta_{\Lambda_b}$ is due to the
presence of the O(r$^2$) term in the potential. Otherwise,
the value of $\beta_{\Lambda}$ is no different than that of B$_s$.
These values are used in the subsequent calculations.
A comment is in order on the choice of the
same wave function for both the baryon as for meson: In the usual procedure, 
the ground state wave function for a baryon is
\begin{equation}
\Psi_{ground} \approx e^{-\alpha^2(r_\lambda^2+r_\rho^2)/2}
\end{equation}
where r$_{\lambda, \rho}$ are the internal coordinates for the three
body system. Due to the idea of considering ${\Lambda_b}$ as
a system containing the bound state of light quarks and a heavy
quark, the separation between the two light quarks which make the
bound state, is treated negligible. Then the
baryon can be considered as a two body system, a reasonable
approximation. The difference between a meson and a baryon
is essentially due to the value of the oscillator strength. 

\subsubsection{Expectation values}
We evaluate the expectation values of the colour-straight operators
only for the vector and axial-vector currents. Nevertheless, the other 
currents can also be studied in the same fashion. Both the currents
are possible for the B-meson, while axial currents vanish for
${\Lambda_b}$ baryon due to the light quarks
constituting a spinless bound state.

Essentially there is no difference between the exponential
ansatz and the harmonic oscillator wave function in representing
the behaviour of the form factor, but they differ while fixing
the scale: in the former case, the hadronic scale of one GeV
is used whereas in the latter the same has been fixed by
solving the Schr\"odinger equation. The two pole anstaz is based
on the well founded experimental values. Basically, the use of the
harmonic oscillator wave function of the constituent quark model
is an alternate picture but in the very same lines of the ansatz.

Hereinafter the operators are referred to by the following
notation: for the meson
\bea
<O_{V, A}^q>_H = \ln \bar b \Gamma_{V, A} b \bar q \Gamma_{V, A} q \rn\\
<T_{V, A}^q>_H = \ln \bar b \Gamma_{V, A} t^a b \bar q \Gamma_{V, A} t^a q \rn
\eea      
where $\Gamma_{V, A} = \gamma_{\mu}, \gamma_{\mu} \gamma_5 $ and 
${t^at^b = 1/2-1/2N_c}$. 
In what follows, q stands for u and d quarks and s for the strange
quark. 

And, ${<O_{V, A}>, <T_{V, A}>}$ will be respectively
denoted as ${\omega_{V, A}, \tau_{V, A}}$. For the baryon, $<O(T)_V>$
correspond to $\lambda_{1, 2}$ respectively.

\subsubsection{B-meson} 
The parameterisation of the matrix element of the colour-straight
operators for the vector current is
\begin{equation}
<B(q) | J_{\Gamma_V} |B(0)> = -v_{\mu} F_B(q^2)
\end{equation}
with the constraints F${_B}$(0) = 1 for valence quark current
and F${_B}$(0) = 0 for the sea quark current. The former is relevant
for the b-meson composition of quarks b${\bar q}$. Then the
corresponding transition amplitude is
\begin{equation}
A_{\Gamma_V}^B = <B(q) | O_V^q  |B(0)> =
v_\mu \int {d^3 q \over {(2 \pi)^3}}F_B(q^2)
\end{equation}
Under isospin SU(2) symmetry,
\bea
\ln O_V^q \rn_{B^-} = 2.88 \times 10^{-3}~\mbox{GeV$^3$}\\
\ln T_V^q \rn_{B^-} = 9.61 \times 10^{-4} ~\mbox{GeV$^3$}
\eea
For B${_s}$, we have 
\bea
\ln O_V^q \rn_{B_s} = 3.83 \times 10^{-3}~\mbox{GeV$^3$}\\
\ln T_V^q \rn_{B_s} = 1.28 \times 10^{-3}~\mbox{GeV$^3$}
\eea
If violation of isospin symmetry and SU(3)
flavour symmetry is considered, then there comes Cabibbo
mixing of eigenstates for d and s quarks among themselves.
That is, $d^{\prime} = d\cos\theta_c+s\sin\theta_c$
and $s^{\prime} = s\cos\theta_c+d\sin\theta_c$. This would bring in
negligible corrections.

For the axial current, there are two form factors which are related 
to each other due to the conservation of the axial current,
${{\partial_\mu} J_{\mu 5}}$ = 0, in the chiral limit.
By the Goldberger-Treiman relation
\cite{don92} which equates axial charge form factor to the coupling 
g${_{B^*B \pi}}$ at q${^2}$ = 0, the operators involving the
axial-currents are estimated in terms of a single form factor.
Thus, given the value of the coupling g, the extraction of the
transition amplitude is similar to the B-meson case.

With the Goldberger-Treiman relation, we obtain the EVFQO for axial currents as
\be
G_1(q^2) = q^2 G_0(q^2) = g e^{-q^2/4\beta^2}
\ee
where g = -- 0.03 \cite{bel95,dos96}. The numerical estimates are
\bea
\ln O_A^q \rn_{B^-} =  8.63 \times 10^{-5}~\mbox{GeV$^3$} \\
\ln T_A^q \rn_{B^-} =  2.88 \times 10^{-5}~\mbox{GeV$^3$} \\
\ln O_A^s \rn_{B_s} =  1.15 \times 10^{-4}~\mbox{GeV$^3$} \\
\ln T_A^s \rn_{B_s} =  3.84 \times 10^{-5}~\mbox{GeV$^3$} 
\eea
These values are insignificant as compared to the vector current contributions.

\subsubsection{$\Lambda_b$ baryon}

For the ${\Lambda_b}$ baryon, treating u and d quarks equally, 
\begin{equation}
\ln O_V^q \rn_{\lbb} = 1.69\times 10^{-2}~\mbox{GeV$^3$}, \hspace{0.6cm}
\ln T_V^q \rn_{\lbb} = 5.64\times 10^{-3}~\mbox{GeV$^3$}
\end{equation}
In the case of ${\Xi_b}$, we have, 
\begin{equation}
\ln \sum_{q^\prime = u,d,s} O_V^{q^\prime} \rn_{\Xi_b} = 2.01 
\times 10^{-2}~\mbox{GeV$^3$}, \hspace{0.6cm}
\ln \sum_{q^\prime = u,d,s} T_V^{q^\prime} \rn_{\Xi_b} = 6.72 
\times 10^{-3}~\mbox{GeV$^3$}
\end{equation}
Even though there are corrections to the form factors due to
charge radius, they are ignored as we are interested in
the wave function density at the origin.

\subsubsection{Non-factorisable part of the FQO}
The nonfactorisable parts of the FQO come in four.
The following is one of the ways of parameterising them
\cite{neub97}.
\bea
{1 \over {2M_B}}<B|(\bar b q)_{V-A} (\bar q b)_{V-A}|B> = 
{\bar f_B^2} M_B B_1/2\\
{1 \over {2M_B}}<B|(\bar bt^aq)_{V-A} (\bar qt^ab)_{V-A}|B> = 
{\bar f_B^2} M_B \epsilon_1/2\\
{1 \over {2M_B}}<B|(\bar b q)_{S-P} (\bar q b)_{S-P}|B> = 
{\bar f_B^2} M_B B_2/2\\
{1 \over {2M_B}}<B|(\bar bt^aq)_{S-P} (\bar qt^ab)_{S-P}|B> = 
{\bar f_B^2} M_B \epsilon_2/2
\eea
where $B_{1, 2}$ and $\epsilon_{1, 2}$ are hadronic parameters.
They are related to $w_{V, A}$ and $\tau_{V, A}$ which are
the expectation values of the operators
$O_{V, A}$ and $T_{V, A}$ as defined earlier:
\bea
{\bar f_B^2} M_B B_1 = \phi_1 = 4(\tau_V+\tau_A)+{2 \over {N_C}}
(\omega_V+\omega_A)\\
{\bar f_B^2} M_B B_2 = \phi_2 = -2(\tau_V-\tau_A)-{1 \over {N_C}}
(\omega_V-\omega_A)\\
{\bar f_B^2} M_B \epsilon_1 = \rho_1 = -{2 \over {N_C}}
(\tau_V+\tau_A)+
(1-{1 \over {N_C^2}})(\omega_V+\omega_A)\\
{\bar f_B^2} M_B \epsilon_2 = \rho_2 = {1 \over {N_C}}
(\tau_V-\tau_A)-
{1 \over {2}}(1-{1 \over {N_C^2}})(\omega_V-\omega_A)
\eea
 
In the case of the $\Lambda_b$ baryon, the nonfactorisable
piece corresponds to
\cite{neub97}
\begin{equation}
<(\bar b q)_{V-A}(\bar q b)_{V-A)} = -{1 \over {2N_C}}
\lambda_1-\lambda_2
\end{equation}

\section{Flavour dependence of decay rates}

In this section, we attempt to evaluate the expectation values of the four-quark
operators from the differences in the total decay rates of the $b$-hadrons. 
That means we assume that the heavy quark expansion, being asymptotic in
nature, converges at $O(1/m^3)$ of the expansion. Already the next-to-leading
order contribution due to terms of $O(1/m^2)$ is only about five percent
of the leading order. Thus, it cannot be anticipated that the size
of the terms at the third order in $1/m$ would be more than a few percent.
On the other hand, the observation made above is applicable only to the 
beauty case,
since
\be
{16 \pi^2 \over {m_c^3}} C(\mu) \left< O_6 \right>_{H_c}   \gg
{16 \pi^2 \over {m_b^3}} C(\mu) \left< O_6 \right>_{H_b}   
\ee
where $C(\mu)$ stands for some structure involving $c_\pm$ and 
$\left< O_6 \right>_H$, the dimension six FQO of hadron.
Hence, in the background described, we make use of the total decay 
rates to obtain the expectation values of the four-quark operators for the
$b$-hadrons. Therefore, the present evaluation depends only
on the heavy quark expansion and the $SU(3)$ flavour symmetry,
as has been shown in \cite{volo}.

\subsubsection{Splitting of Decay Rates and Expectation Values of 
Four-quark Operators}
The $B$ mesons, $B^-$, $B^0$ and $B^0_s$, are triplet
under $SU(3)_f$ flavour symmetry. The total decay rate of $B$ meson
splits up due to its light quark flavour dependence at the third order
in the HQE. The differences in the decay rates of the triplet,
$\gm(B^0) - \gm(B^-), \gm(B^0_s) - \gm(B^-)$ and
$\gm(B_s^0) - \gm(B^0)$, are related to the third order terms
in $1/m$ by
\bea
\Gamma(B^0)-\Gamma(B^-) &=& -\Gamma^\prime_0 (1-x)^2
\left\{Z_1{1 \over 3}(2c_+ - c_-+6)+2(c_+-c_-/2+1)\right\} \nonumber\\
&&\times \left<O_6\right>_{B^0-B^-} \label{one}\\
\Gamma(B^0) - \Gamma(B^-) &=& -\Gamma^\prime_0 (1-x)^2
\left\{Z_2{1 \over 3}(2c_+ - c_-+6)+2(c_+ - c_-/2+1)\right\} \nonumber\\
&&\times \left<O_6\right>_{B^0_s-B^-}\\
\Gamma(B^0)-\Gamma(B^0) &=& -\Gamma^\prime_0 (1-x)^2
\left\{(Z_1-Z_2){1 \over 3}(2c_+-c_-+6)\right\} \nonumber\\
&&\times \left<O_6\right>_{B^0_s-B^0} \label{three}
\eea
where $\Gamma^\prime_0$ = $2G_f^2|V_{cb}|^2m_b^2/3\pi$,
$c_- = c_+^{-2} = (\alpha(m)/\alpha(m_W))^{2\gamma/\beta}, 
\beta = 11-2n_f/3$ ($n_f$ the number of flavours), $\gamma =2$, 
$x$ = $m_c^2/m_b^2$ and
\bea
Z_1 &=& \left(\cos^2\theta_c(1+{x \over 2})+
\sin^2\theta_c \sqrt{1-4x}(1-x)\right)\\
Z_2 &=& \left(\sin^2\theta_c(1+{x \over 2})+
\cos^2\theta_c \sqrt{1-4x}(1-x)\right)\\
\left<O_6\right> &=& \left<{1 \over 2}(\bar b \Gamma_{\mu} b)
[(\bar d \Gamma_{\mu} d)
-(\bar u \Gamma_{\mu} u)]\right> \nonumber\\
&&\equiv 
\left<{1 \over 2}(\bar b \Gamma_{\mu} b)[(\bar s \Gamma_{\mu} s)
-(\bar q \Gamma_{\mu} q)]\right>
\eea
with $q = u, d$.
In (\ref{one}-\ref{three}), 
the right hand side contains the terms corresponding to the
unsuppresed and suppressed nonleptonic decay rates and
twice the semileptonic decay rates at the third order.

For the decay rates $\gm(B^-) = 0.617$ ps$^{-1}$,     
$\gm(B^0) = 0.637$ ps$^{-1}$ and $\gm(B^0_s) = 0.645$ ps$^{-1}$ \cite{pdg},
the EVFQO are obtained for a $B$ meson, as an average from
(\ref{one}-\ref{three}):
\be
\left<O_6\right>_B = 8.08 \times 10^{-3} ~\mbox{GeV$^3$}.
\label{bval}
\ee
This is smaller than the one obtained
in terms of the leptonic decay constant, $f_B$.

On the other hand, for the triplet baryons,
$\Lambda_b$, $\Xi^-$ and $\Xi^0$, with
$\tau(\Lambda_b)$ $<$ $\tau(\Xi^0) \approx \tau(\Xi^-)$,
we have the relation between
the difference in the total decay rates and the terms of
$O(1/m^3)$ in the HQE, as
\be
\Gamma(\Lambda_b)-\Gamma(\Xi^0) = {3 \over 8}\Gamma_0^\prime (-c_+(2c_-+c_+-2)
\left<O_6\right>_{\Lambda_b-\Xi^0}
\ee
We obtain the EVFQO for
the baryon
\be
\left<O_6\right>_{\Lambda_b-\Xi^0} = 3.072 \times 10^{-2} ~\mbox{GeV$^3$}
\label{lval}
\ee
where we have used the
decay rates corresponding to the lifetimes 1.24 ps and 1.39 ps \cite{pdg}
of $\Lambda_b$ and $\Xi^0$ respectively.
The EVFQO 
for the baryon is about 3.8 times larger than that of the B meson. But it is about
3.5 - 4.0, depending on the 
changes in the mass and other sources of uncertainty.

The expectation values for a baryon is quite large. In fact, Rosner \cite{rosner96}
found it to be 1.8 times larger than that of a meson which accounts
for about 45\% (if hard gluon exchange is included)
of the needed enhancement in the decay rate. However, what we obtain
is model independent. We have only used the experimental total decay
rates. The result is surprising as well as genuine, if one has to believe
the experimental values used. Still, there is room for uncertainty
of a few percent. On the other hand, the generic structure we employed
can be decomposed into many as found in \cite{neub97}
by Neubert and Sachrajda. Using them. we will get an improved estimate,
but in that case it may become somewhat model dependent. We should note here
that preliminary study on lattice \cite{sach} shows that the EVFQO would
enhance the decay rate of $\lbb$.

\chapter{Spectator effects}
In this Chapter, we present the predictions for the ratio of lifetimes of beauty hadrons using
the expectation values obtained in Ref. \cite{arun1,arun2}. Also presented
is the determination 
the inclusive charmless semileptonic decay rate of $\lbb$ and discuss its
influence on $Br(b \ra X_u l \nu)~$\footnote{S. Arunagiri, {\it Inclusive charmless semileptonic decay of
$\lbb$ and $Br(b \ra X_u l \nu)$}, hep-ph/0005266. \cite{arun3}}.

\section{Spectator Processes}

\begin{figure}
 \begin{center}
 \epsfig{figure=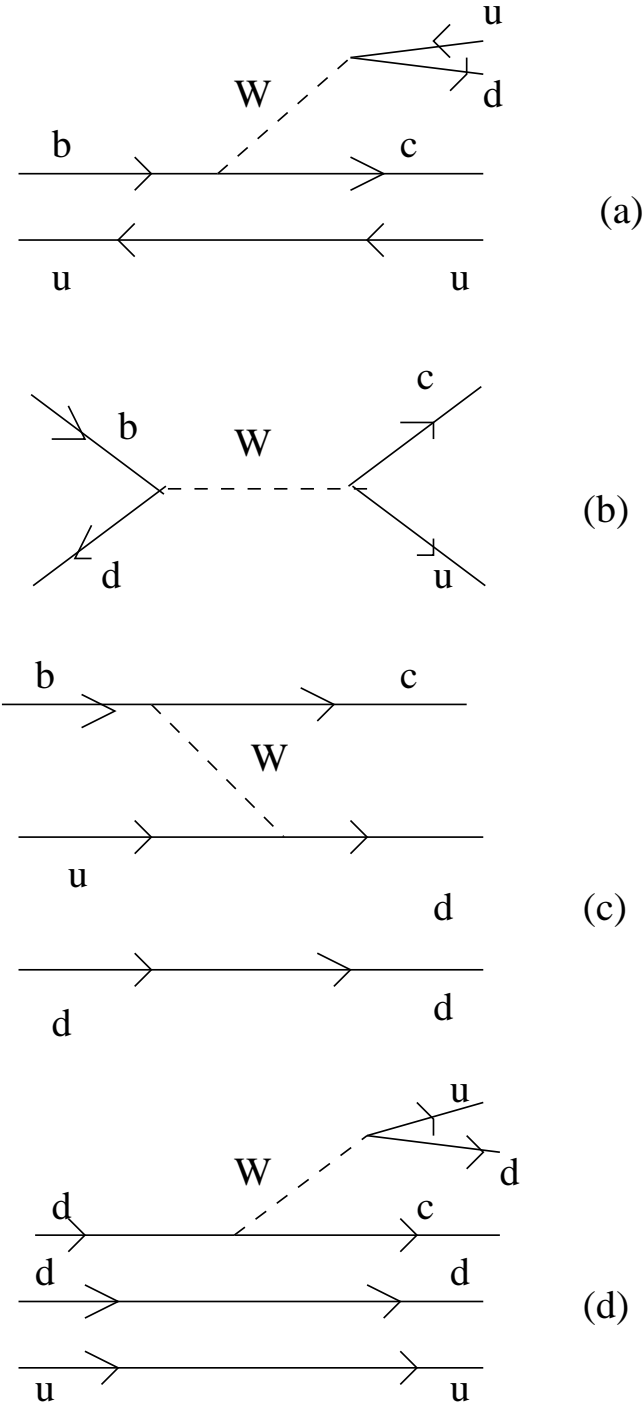}
 \caption{Spectator quark processes. (a) Pauli interference in B,
 (b) Weak annihilation in B, (c) W-scattering in $\lbb$
  and (d) Pauli interference in $\lbb$.}
 \label{fig:3.14}
 \end{center}
\end{figure}

In the description of the inclusive decays of heavy hadrons by
the heavy quark expansion, the total decay rate scales like $m_b^5$.
Corrections appear at $O(1/m_b^2)$ due to the heavy quark motion
inside the hadron and the hyperfine interaction due to the heavy quark
spin orientation. These corrections are about 10\% of the leading order.
As the heavy quark mass increases, the corrections are much
suppressed by the inverse powers of the heavy quark mass. At the third
order in $1/m_b^3$, the effects are due to the participation of the
light quarks (See Figure\ref{fig:3.14}). Though the effects are negligibly small, they
distinguish the lifetimes of various hadrons of given flavour
quantum number. We discuss first the processes involving light quarks 
\cite{gnb,bgt,brt,sv}.
Then we will use the expectation values of the four-quark operators
of mesons and baryons estimated in the  previous chapter to get
the lifetimes of the beauty hadrons.

\subsubsection{Pauli interference (PI)}
For the decay of a $(Q \bar q)$ hadron, the quark level process is
$Q(q_i) \ra Q^\prime q_f q_f^\prime (q_i)$, where the subscript associated
with $q$'s denote the light quark in the initial hadron $i$ and 
the light quarks produced in the decay $f$. If one of the final state
light
quarks happens to be the same as the initial state quark, then
the wave functions will interfere. 
For example consider the decay mode
$B^- \ra X_c X_q$, where the subscripts $c$ and $q$
denote the charmed and light quark of either $u$ and $d$ final states. If either
of the light quarks in the final state $X_q$ is identical to the
one in the initial hadron, they will interfere destructively.
On the other hand, if $Q^\prime$ happens to be identical to $q_i$, then
the
interference will be a constructive one. The former will decrease the decay
rate
causing increase in the lifetime, while the latter will do the opposite.

The constructive interference can happen only in the baryon. For example,
in the charmless semileptonic processes of $\Lambda_b \ra X_u l \nu$,
the corresponding quark process is $b(ud) \ra X_u l \nu (ud)$. The
$u$ quark in the state $X_u$ and the $u$ in the
initial hadron will interfere constructively. This can happen in the
nonleptonic process as well.

\subsubsection{$W$-scattering (WS) and weak annihilation (WA)}
Another spectator process corresponds to the cross-channel
of the quark decay. In the notation used before, the light quark in the
initial hadron can scatter off the heavy quark through exchange of
$W$ field into a pair of light quarks, {\it i. e.,}
$q_i Q \ra q_f q_f^\prime$. This process can happen only in baryons.
This process is known as $W$-scattering or $W$-exchange, which occur in
baryons only.

Or, if the quark scattering off the heavy one happens to be an anti-quark,
then they annihilate into a pair of leptons or quarks via a $W$ boson:
$q_i Q \ra q_f~q_f^\prime$, where $q_i$ is an anti-quark. This
is the so-called weak annihilation process occuring only in mesons.
\section{Decay Rates due to Spectator Effects}
We summarise below the spectator contribution to the beauty
hadron decays.

The quark level process $b \ra c \bar u d$:
\begin{itemize}
\item
$B^- (b \bar u)$: the $\bar u \bar u$ interference
term is given by
\be
\Delta\Gamma(B^-) = \Gamma_0 24 \pi^2 C_0{<O_6^q>_{B^-}/{m_b^3}} \label{ddbb}
\ee
where 
\bea
C{_0} &=& {c_+^2-c_-^2+(c_+^2+c_-^2)/N_c}\\
{\Gamma_0} &=& G_f^2 m_b^5 |V_{KM}|^2 / 192 \pi^3\\
c_+ = c_-^{-1/2} &=& (\alpha(m_b)/\alpha(m_W))^{4/\beta_0} 
\eea
with $\beta_0 = 11-2n_f/3, n_f$ being
the number flavours. In (\ref{ddbb}), $\left<O_6\right>$
stands for the EVFQO.

\item
$B^0_d (b \bar d)$: In this case, the
$d \bar d$ annihilation contributes to the total decay rate. The
magnitude of this effect is not significant.

\item
$B_s^0 (b \bar s)$: No spectator effect occurs.

\end{itemize}

The spectator effects become significant depending on the quark level
process.

For the above quark level process, in the case of beauty baryons,
we have:
\begin{itemize}
\item
$\Lambda_b (bud)$: There are two processes, $dd$ destructive Pauli interference
and $u \bar u$ W-scattering. The Pauli interference in this
case enhances the decay rate resulting in the $\Lambda_b$ lifetime
smaller than the average lifetime of the beauty hadrons. The combined
decay rates are given in (\ref{wsb} and \ref{pib}) in the latter part
of this Chapter.

\item
$\Xi^0 (bds), \Xi^- (bus)$: In the case of these hyperons, weak scattering
and destructible Pauli interference occur.
\end{itemize}

\section{Lifetimes of Beauty Hadrons}
Using the results obtained for the expectation values of the
four-quark operators of the hadrons, we present the results for
the ratio of lifetimes of them.

The decay rates of the b-flavoured hadrons are given by

\begin{equation}
\Gamma(H_b \rightarrow H_c l \bar \nu_l) = \Gamma_0
\left [(1-{\lambda_1-3\lambda_2 \over {2m_Q^2}})f(x)+
{2\lambda_2 \over {m_Q^2}}f'(x)\right ]   \label{ratee}
\end{equation}

where, with $x = m_c^2/m_b^2$
\bea
f(x) &=& 1-8x+8x^3-x^4+12x^2 \mbox{ln} x^2 \\
f'(x) &=& (1-x)^4
\eea
are the QCD phase space factors and $\lambda_1$ and $\lambda_2$  
correspond to the motion of the heavy quark inside the hadron and 
the chromomagnetic interaction respectively. We use:
$\lambda_1(B) = 0.5$ GeV$^2$, $\lambda_1(\lbb) = -0.43$ GeV$^2$ and 
$\lambda_2(B) = 0.12$ GeV$^2$.
Equation (\ref{ratee}) is further supplemented by the FQO at 
the order (1/m$_Q^3$) in the HQE. 

\subsection{Ratio of Lifetimes of B$^-$ and B$_d$}
Although the difference between the lifetimes of the charged and
the neutral B-mesons is almost a settled issue, we check them once
again using the expectation values of the colour-straight operators.
This difference is attributed to PI and WA. Neglecting the WA as
it is strongly CKM suppressed the result for  the PI is
\begin{equation}
\Delta\Gamma^f(B^-) = \Gamma_0 24 \pi^2 C_0{<O_V^q>_{B^-}-
<O_A^q>_{B^-} \over {m_b^3}}
\end{equation}
where C${_0}$ = ${c_+^2-c_-^2+{1 \over {N_c}}(c_+^2+c_-^2)}$ and 
the values for Wilson coefficients are: c${_+}$ = 0.84 and c${_{--}}$
= -1.42 with N${_c}$ = 3. We use m${_b}$ = 4.8 GeV and
${|V_{cb}|}$ = 0.04. Then the ratio is
\begin{equation}
{\tau(B^-)\over {\tau(B_d)}} = 1.00  
\end{equation}
This agrees well with the one obtained in terms of  B-meson decay
constants.

The decay rates due to spectator quark(s) processes are:
For B$^-$,
\begin{equation}
\Delta\Gamma^{nf}(B^-) = {G_f^2 m_b^2 \over {12\pi}}
|V_{cb}|^2(1-x)^2
[(2c_+^2-c_-^2)\phi_1+3(c_+^2+c_-^2)\rho_1]
\end{equation}
Hence the ratio is 1.03.

\subsection{Ratio of Lifetimes of B$^-$ and B$_s$}
The difference in lifetimes between the two neutral mesons
B$_s$ and B$_d$ is
due to $W$ exchange. The numerical result is 
\begin{equation}
{\tau(B_s) \over {\tau(B_d)}} = 1.00  
\end{equation}

Corresponding to the nonfactorisable part, we get the decay rate
\be
\Delta\Gamma^{nf}(B_s) = {\overline \Gamma_0} (1-4x)^{1/2}
 \eta
\ee
where $\overline \Gamma_0 =
{G_f^2 m_b^2/{12\pi}}|V_{cb}|^2$ and $\eta$ is given by
\be
\eta = 
{(2c_+-c_-)^2 \over {3}}\left[((1-x)\phi_{1}^s-(1+2x)\phi_{2}^s)+
{(c_++c_-)^2 \over {2}}((1-x)\rho_{1}^s-(1+2x)\rho_{2}^s)\right]
\ee
The ratio becomes 1.02.

\subsection{Ratio of Lifetime of ${\Lambda_b}$ and B$^-$}
In the HQE, the difference in lifetimes between mesons and
baryons begins to appear at order 1/m$_Q^2$. Nevertheless, it
is dominant at third power in 1/m$_Q$. At this order, the
FQO receives corrections due to WS and PI.
They are
\begin{eqnarray}
\Gamma_{WS}(\Lambda_b) = 92 \pi^2 \Gamma_0c_-^2
{<O_V^q>_{\Lambda_b}
\over {m_b^3}} \label{wsb}\\
\Gamma_{PI}(\Lambda_b) = -48 \pi^2 \Gamma_0C_1
{<O_V^q>_{\Lambda_b}
\over {m_b^3}} \label{pib}
\end{eqnarray}
where $C_1 = -c_+(2c_--c_+)$. As mentioned earlier, PI is
destructible for radiative corrections and it enhances the
decay rate leading to smaller lifetime for ${\Lambda_b}$.
The effect of WS, on the other hand, is colour enhanced
and its consequence is smaller. Hence, 
\begin{equation}
{\tau(\Lambda_b) \over {\tau(B^d)}} = 0.79
\end{equation}

The decay rate modified by the nonfactorisable piece is given by 
\begin{equation}
\Delta \Gamma (\Lambda_b) = {G_f^2 m_b^2 \over {16 \pi}}
\bar \lambda
[4(1-x)^2(c_-^2-c_+^2)-(1-x)^2(1+x)(c_--c_+)(5c_+-c_-)]
\end{equation}
Correspondingly,
the ratio
is
\begin{equation}
{\tau(\Lambda_b) \over {\tau(B^-)}} = 0.84
\end{equation}
In mesonic cases, the nonfactorisable piece gives a little
higher  value. In particular, the ratio of
the lifetimes of the baryon and meson is significantly larger.  

\section{Inclusive Charmless Semileptonic Decay}
\subsubsection{Motivation}
Precise determination of the CKM matrix elements is fundamentally
an important task for the Standard Model. There are as many as
five of the nine elements of the CKM matrix that can be extracted
from the knowledge of the weak decays of beauty hadrons.
Particularly, $|V_{cb}|$ and $|V_{ub}|$ can be extracted in the framework
of heavy quark expansion (HQE) in a direct and fairly model independent way
\cite{neub0}.
With the measurements of the inclusive charmless semileptonic branching ratio of $b$
hadrons by the ALEPH \cite{aleph}, L3 \cite{L3} and DELPHI \cite{d}
collaborations at LEP, the
determination of $|V_{ub}|$ has attracted renewed interest recently.
The measured values of inclusive charmless semileptonic branching ratio of 
$b$ hadrons and the corresponding
values of $|V_{ub}|$ extracted are
\bea
Br(b \rightarrow X_u l\nu_l) &=& (1.73 \pm 0.55 \pm 0.55) \times 10^{-3},\nonumber\\
    |V_{ub}| &=& (4.16 \pm 1.02)\times 10^{-3}\hspace{0.2in} ~\mbox{(ALEPH)} \\
Br(b \rightarrow X_u l\nu_l) &=& (3.3 \pm 1.0 \pm 1.7) \times 10^{-3},\nonumber\\
   |V_{ub}| &=& (6.0^{+0.8+1.4}_{-1.0-1.9} \pm 0.2) \times 10^{-3}.
   \hspace{0.7in} ~\mbox{(L3)} \\
Br(b \rightarrow X_u l\nu_l) &=& (1.57 \pm 0.35 \pm 0.48 \pm 0.27)\times10^{-3},\nonumber\\
   |{V_{ub}/ {V_{cb}}}| &=& (0.103^{+0.011}_{-0.012} \pm 0.016 \pm 0.010).
    \hspace{0.1in} ~\mbox{(DELPHI)}
\eea
In these analyses, the measured quantity is
\be
{Br(b \rightarrow u e \nu) \over {Br(b \rightarrow c e \nu)}}
\propto \left| {V_{ub} \over {V_{cb}}} \right|^2\,.
\ee
This has an advantage of being free of many hadronic uncertainties
that occur in the non-leptonic decays. Also,
the determination of the CKM matrix elements from inclusive decays can in general be made
with less theoretical uncertainties than the ones extracted from exclusive modes.
For semileptonic decays of baryons, however, there could still
be large spectator effects due to Pauli interference as pointed out
by Voloshin for $\Xi_c$~\cite{voloshin}.
To the inclusive charmless semileptonic branching ratio of
$b$ hadrons, the contribution of $\Lambda_b$ is about 10\%
with the rest coming from the $B$ mesons. There are many theoretical works
\cite{jin} which study the $B \rightarrow X_u l \nu_l$ to determine
$|V_{ub}|$. But not much has
been done on the inclusive charmless semileptonic decay of
$\Lambda_b$. In the analyses of ALEPH and L3, 
$\Lambda_b \rightarrow X_u l \nu_l$ is included in their sample,
but in that of DELPHI, the rejection of kaons and protons 
that was used in the selection criteria
reduces the contributions from $\Lambda_b$.

We study the inclusive charmless semileptonic decay of
$\Lambda_b$ \cite{arun3}. We find that the baryonic decay rate is larger
by a factor of about 1.36, due to spectator effects, than that of
the $b$ quark decay rate. We then discuss the correction factors
needed on $Br(b \rightarrow u l \nu_l)$ to account for the
spectator effect in $\Lambda_b \rightarrow X_u l \nu_l$.

\subsection{Inclusive Charmless Semileptonic Decay Rates}
The total rate of inclusively decaying beauty hadrons into a
charmless final state is given by the HQE\cite{neub0} as
\be
\Gamma(H_b) = {G_f^2 |V_{ub}|^2 m^5 \over {192 \pi^3}}
\left( 1 - {\lambda_1 - 3 \lambda_2 \over {2m^2}} - 2 {\lambda_2
\over {m^2}} + O\left(1 \over {m^3}\right)\right)
 \label{rate}
\ee
where, with $m$ being the mass of the heavy quark, the term of
$O(1/m^0)$ corresponds to the free heavy quark
decay rate, the terms at $O(1/m^2)$ describe the motion of the
heavy quark inside the hadron ($\lambda_1$ = $-$0.5  GeV$^2$ for $B$ and
$-$0.43  GeV$^2$ for $\Lambda_b$) and the chromomagnetic
interaction due to the heavy quark spin projection
($\lambda_2$ = 0.12  GeV$^2$)
which vanishes for baryons except $\Omega_Q$ and
the third order term in $1/m$ is given by
\begin{center}
$C(\mu)\left<H|(\bar b \Gamma q)(\bar q \Gamma b)|H\right>$.
\end{center}
The operators in $\left<...\right>$, denoted as $\left<O_6\right>$ below,
are evaluated in \cite{arun}
by one of us for $B$ and $\Lambda_b$. The Wilson coefficients
are describing the spectator quark processes like Pauli
interference, weak annihilation and $W$-scattering
which occurs in baryons only.

In the mode $\Lambda_b \rightarrow X_u l \nu_l$, the
spectator effect is the constructive interference of the $u$ quark
of the final state with the $u$ quark in the initial hadron 
(see Fig.~\ref{fg:diag}) which
enhances considerably the decay rate. Otherwise,
without spectator effects, the baryonic decay rate is almost the same
as that of the heavy quark decay rate.
The decay rate due to spectator processes is given by
\be
 {\Delta \Gamma (\Lambda_b) = 48 \pi^2 \Gamma_0 
 {\left<O_6\right>_{\Lambda_b} \over {m^3}}}
\ee
where $\Gamma_0 = G_f^2 |V_{ub}|^2 m^5/192 \pi^3$. Using the
expectation values of the four-quark operators obtained in Ref. \cite{arun}
for $\Lambda_b$, $\left<O_6 \right>_{\Lambda_b} = 6.75 \times 10^{-2}$ GeV$^3$,
we obtain the ratio
\be
  {{\Gamma(\Lambda_b \rightarrow X_u l \nu_l)} \over
  {\Gamma(b \rightarrow X_u l \nu_l)}} = 1.36.
  \label{ratio}
\ee
whereas at $O(1/m^2)$, this ratio is
1.01. It should be pointed out that
this enhancement has a large theoretical uncertainty which is difficult to estimate.
On the other hand, no such enhancement exists for 
$\Lambda \rightarrow X_c e \nu_e$ and thus the ratio 
is increased by 36\% for $\Lambda_b$. 
In passing, we note that
the hadronic $b \rightarrow u$
decay of $\Lambda_b$ is not substantially enhanced due to the 
cancellations of constructive $uu$ and destructive $dd$ interferences. 

\subsection{$Br(b \ra X_u l \nu)$ and $V_{ub}$}
The inclusive charmless semileptonic branching ratio of $b$ hadrons is
given by 
\be
Br(b \rightarrow X_u l \nu_l) = (1-f_{\Lambda_b}) 
\Gamma(B \rightarrow X_u l \nu_l) \tau(B) + f_{\Lambda_b}
\Gamma(\Lambda_b \rightarrow X_u l \nu_l)\tau(\Lambda_b)
\ee
In principle, $\Xi_b^{0,-1}$ can also be produced at LEP. However,
it is suppressed by $s\bar s$ production from vacuum relative to
$u\bar u$ or $d\bar d$ production. Furthermore, the neutral $\Xi_b$
which consists of $b$, $u$, and $s$ quarks has the same enhancement
factor as $\Lambda_b$ in the charmless semileptonic decay rate, even though
the charged $\Xi_b$ which is made of $bds$ quarks
does not have the corresponding enhancement.
We will thus assume in this analysis that all weakly decaying b-baryons
are $\Lambda_b$'s.
Thus using the above estimate with $f_{\Lambda_b}$ = 10\%,
$\tau(B)$ = 1.64 ps and $\tau(\Lambda_b)$ = 1.24 ps, we obtain
the following value for inclusive charmless semileptonic branching ratio of $b$ hadrons:
\be
Br(b \rightarrow X_u l \nu_l)  =  1.16 \times 10^{-3}
\ee
If the spectator effects are
not included, then
\be
 Br(b \rightarrow X_u l \nu_l)  =  1.13 \times 10^{-3}
 \label{brbunsp}
\ee
In the above estimate in Eqs. (\ref{ratio}-\ref{brbunsp}), we have employed
$|V_{ub}| = 3.3 \times 10^{-3}$ and $m$ = 4.5 GeV.
Also, we have used the decay rate for
$B$ mesons as estimated at $O(1/m^2)$ using the Eq. (\ref{rate}),
$Br(B \rightarrow X_u l \nu_l) = 1.155 \times 10^{-3}$.
No spectator effect occurs in $b \rightarrow u l \nu$ transition in $B$
mesons except the negligible weak annihilation in $B^-$.
Also, the experimental selection criteria used for  
the $X_u$ system mostly reject such final states.

The corrections on $|V_{ub}|$ due to the enhancement of 
$\Lambda_b \rightarrow X_u l \nu_l$ is about $-$1.3\%, and
with these corrections, the central values for $|V_{ub}|$ by
ALEPH and L3 become, as given by
$|V_{ub}| = \left({Br_{expt} / {Br_{theory}}}\right)^{1/2}0.0033$,
as

\[ |V_{ub}|  = \left\{
\begin{array}{ll}
 4.03 \times 10^{-3} & \mbox{(ALEPH)}\\
 5.57 \times 10^{-3} & \mbox{(L3)}
\end{array}
\right. \]

Finally, we observe that
(1) the contribution of $\Lambda_b$ definitely influences
the branching ratio due to a constructive
Pauli interference and (2) such spectator effects in $\Lambda_b$
decay have to be taken into account for a precise
determination of $|V_{ub}|$ which relies on charmless semileptonic
decays.

\chapter{Short distance nonperturbative corrections}
In this Chapter, we discuss the power corrections to the parton decay
rate, due to renormalons, of the heavy hadron decays \footnote{
S. Arunagiri, On the short distance nonperturbative corrections
in heavy quark expansion, hep-ph/0009109 \cite{arun}.}. 
\section{Renormalons}
As early as 1952, Dyson \cite{fjd} showed that the perturbative expansion in
a renormalisable theory like QED is not convergent. It is also true
in QCD. This issue poses an important conceptual problem of giving a
meaning to perturbation theory. Dyson's arguement is the following.
An observable in QED, $F(e^2)$ is expressed in an expansion in $e^2$
as
\be
F(e^2) = f_1 e^2 + f_2 e^4 + \cdots
\ee
If there is a radius of convergence around $e^2 = 0$,
then there will be a convergent result also for $-e^2$.
The latter corresponds to an attractive force of equal charges
and a repulsive force of opposite charges. This would physically
mean that the state corresponding to the equal charges has unbounded
negative energy. Thus the perturbative series cannot converge for 
negative $e^2$. Hence the radius of convergence is zero.

The divergent series can be dealt by the Borel summation technique
\cite{gth,bene}.
A divergent perturbation series
\be
f[a] = \sum_{n = 1}^\infty f_n a^n = f_1a+f_2a^2+\cdots
\ee
can be re-expressed by the Borel transform $B[b]$.
\be
B[b] = \sum_{n = 0}^\infty f_{n+1} {b^n \over {n!}} \approx
f_1+f_2b+f_3 {b^2 \over 2}+\cdots
\ee
The original series $F[a]$ can be reproduced by
\be
f[a] = \int_0^\infty db ~e^{-b/a} B[b]
\ee
This would give a meaning to the series $f[a]$ by the fact that the integral
converges and $B[b]$ has no singularities in the integration range.

\begin{figure}
 \begin{center}
 \epsfig{figure=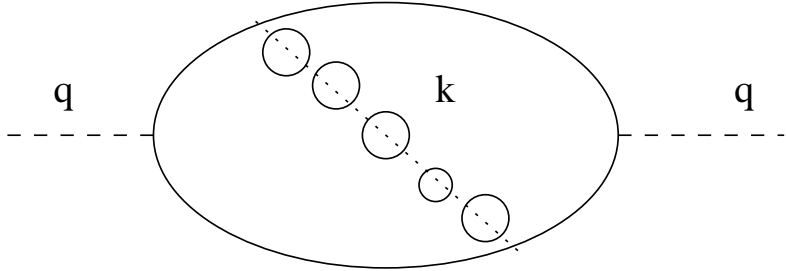}
 \caption{Renormalon graph (dotted lines are gluons).}
  \label{fig:4.1}
 \end{center}
\end{figure}

To be more explicit, let us consider
\be
\int d^4x ~e^{iqx} \ln 0|T \{j_\mu(x)j_\nu(0)\}|0 \rn
=(g_{\mu \nu}q^2 - q_\mu q_\nu) \Pi(q^2)
\ee
With $s = q^2$, the Adler function is defined as
\be
D(s) = -s {d \over {ds}} \Pi(s)
\ee
For the case of QED or QCD, we consider
\be
D_n(Q^2) = \int d^4k ~F(k^2, kq, Q^2) {a \over {k^2}} (blob)^n
\ee
where we defined $a = \alpha(Q62)/\pi$ and $(blob)^n$ represents
$n$ loops in a gluons line (Figure \ref{fig:4.1}). 
Then the above equation would have the form
\be
D_n(Q^2) = \int dk^2 ~k^2 \Phi \left({k^2 \over {Q^2}}\right)
{q \over {k^2}} \left[\beta a\log \left({k^2 \over {Q^2}}\right)\right]^n
\ee
where $\beta$ is the first coefficient of the beta function. Finally,
for momenta above and below $Q^2$, we have the expressions
corresponding respectively to the ultraviolet and infrared regions:
\bea
D(Q^2)_{UV} = {Q^2 \over {n+1}} \int_0^\infty db~ e^{-b/a} {1 \over {\beta
{b \over {n+1}}}}\label{ir}\\
D(Q^2)_{IR} = {Q^2 \over {m-1}} \int_0^\infty db ~e^{-b/a}
{1 \over {\beta
{b\over {m-1}}}}\label{uv}
\eea

where $b = a \log (Q^2/k^2)(h +1)$ with $h = m,n$.

In the infra red case, (\ref{ir}), the poles in the Borel plane
correspond to
$b = -(n+1)/\beta$ on the negative axis. Hence the poles are not
in the integration range. Thus the poles are at $b = -2/\beta, -3/\beta,
\cdots$. For the ultraviolet case, (\ref{uv}), the poles are at
$b = +1/\beta, +2/\beta, \cdots$. In this case, $b = (m-1)a \log (k^2/Q^2)$.
Thus, in this case, poles are in the range of integration and hence they
are Borel summable.

In the case of QCD which is asymptotically free in the UV region, the
IR and UV renormalons defined above are interchanged. That is, the
UV poles are at $b = -1/\beta, -2/\beta, \cdots$ and the IR poles
at $b = +2/\beta, +3/\beta, \cdots$.

\subsubsection{Operator Product Expansion}

The divergence of perturbation theory is deeply connected with
the operator product expansion (OPE). Consider the OPE
\be
D(Q^2) = [\mbox{parton model}](1+a_1 \alpha(Q^2) + a_2 \alpha(Q^2)^2 + \cdots)
+ \sum_{n = 4, 6, ...} C_n\left<O_n\right>
\ee
where $C$ stands for the coefficient functions, $\left<O_n\right>$
the operators of dimension $n$. We cannot have a dimension two operators.
So the term $O(1/Q^2)$ is absent. The perturbative part of the OPE
receives the renormalon corrections \cite{zak}.
It has been shown that the absence of
the first power suppressed term corresponds to the absence of singularity
at
$b = +1/\beta$ in the Borel plane. However, the renormalon
corrections to the perturbative series have one-to-one correspondence
with the power suppressed terms.

\section{Power Corrections}

The divergence of the perturbation theory at large order brings in
an ambiguity to physical quantities specified at short distances.
According to the present understanding, the ambiguity is given
by a class of renormalon diagrams which are chain of
$n$-loops in a gluon line. The phenomenon is deeply connected with
the operator product expansion (OPE). The perturbative part of the OPE
receives the renormalon corrections \cite{zak}.
Since in the OPE the first power-suppressed nonperturbative term is absent
and the renormalon corrections constitute short distance
nonperturbative effect, 
they are more significant than large order
corrections. 

The phenomenology of the power corrections is significant
for the heavy quark expansion (HQE) to describe the
inclusive decays of heavy hadrons by an expansion in the
inverse powers of the heavy quark mass, $m_Q$.
As the inclusive decay rate of heavy hadrons scales like
the fifth power of the heavy quark mass,
the power corrections arise due to momenta smaller than the
heavy quark mass. However, these IR renormalons would,
being nonperturbative effect, have greater influence in
the HQE prediction of quantities of interest.
These short distance nonperturbative effects
can be sought for explaining the smaller lifetime of $\Lambda_b$.
We should note that these power corrections to heavy quark decay rate
represents the breakdown of the quark-hadron duality. Therefore, it may shed
light on the working of the assumption of quark-hadron duality in
the heavy quark expansion.

In this section, we study the renormalon corrections considering the
heavy-light correlator in the QCD sum rules approach, assuming that the
nonperturbative short distance corrections given by the gluon mass are much
larger than the QCD scale. We carry out the analysis for both heavy meson
and a heavy baryon. Our study shows that the short distance
nonperturbative corrections to the baryon and the meson
differ by a small amount which is significant for the smaller lifetime of the $\Lambda_b$.

Let us consider the
correlator of hadronic currents $J$:
\be
\Pi(Q^2) = i \int d^4x e^{iqx} \ln 0|T\{J(x)J(0)\}|0\rn
\ee
where $Q^2 = -q^2$. The standard OPE is expressed as
\be
\Pi(Q^2) \approx [~\mbox{parton model}](1+a_1 \al + a_2 \al^2 + ....)
+ O(1/Q^4)
\ee
where the power suppressed terms are quark and gluon operators.
The perturbative series in the above equation can be rewritten as
\be
D(\al) = 1 + a_0 \al + \sum_{n = 1}^\infty a_n \al^n \label{ddd}
\ee
where the term in the sum is considered to be the
nonperturbative short distance quantity.
It is studied by Chetyrkin {et al} \cite{chet} assuming that the short
distance tachyonic gluon mass, $\lambda^2$,
imitates the nonperturbative physics of the QCD. This, for the gluon
propagator, means:
\be
D_{\mu \nu}(k^2) = {\delta_{\mu \nu} \over {k^2}} \ra
\delta_{\mu \mu}\left( {1 \over {k^2}}+{\lambda^2 \over {k^4}}\right)
\ee
The nonperturbative short distance corrections are argued to be the
$1/Q^2$ correction in the OPE.

Let us consider the assumption of
the gluon mass $\lambda^2 \gg \lq^2$ which
is not necessarily to be tachyonic one.
The feature of the assumption can be seen
with the heavy quark potential
\be
V(r) = -{4\alpha(r) \over {3r}}+kr \label{v}
\ee
where $k \approx$ 0.2 GeV$^2$,
representing the string tension. It has been argued in
\cite{bal} that the linear term can be replaced by a term of order $r^2$.
It is equivalent to replace $k$ by a term describing the ultraviolet
region.
For the potential in (\ref{v}),
\be
k \ra constant \times \al \lambda^2
\ee
In replacing the coefficient of the term of $O(r)$ by $\lambda^2$,
we make it consistent by the renormalisation factor.
Thus the coefficient $\sigma(\lambda^2)$ is given by \cite{ani}:
\be
\sigma(\lambda^2) = \sigma(k^2)\left(\alpha(\lambda^2) \over
\alpha(k^2)\right)^{18/11} \label{rge}
\ee
Introduction of $\lambda^2$ brings in a small correction to the Coulombic term.
By use of (\ref{rge}), we specify the effect at both the ultraviolet
region
and the region characterised by the QCD scale.
Then, we rewrite (\ref{ddd}) as
\be
D(\al) = 1 + a_0 \al\left(1+{k^2 \over {\tau^2}}\right)
\label{dc}
\ee
where $\tau$ is some scale relevant to the problem and $k^2$ should be
read
from (\ref{rge}). We would apply this to
the heavy light correlator in heavy quark effective theory.

We should note that in the QCD sum rules approach, the scale involved
in is given by the Borel variable which is about 0.5 GeV. But
in the heavy quark expansion the relevant scale is
the heavy quark mass, greater than the
hadronic scale. Thus, there it turns out to be infrared renormalons
effects.
But, still it represents the short distance nonperturbative
property, by virtue of the gluon mass being as high as the hadronic
scale.

{\it Meson:} For the heavy light current, $J(x) = \bar Q(x) i\gamma_5
q(x)$,
the QCD sum rules is already known \cite{ben}:
\be
{\tilde f}_B^2 e^{-(\bl)} = {3 \over {\pi^2}}
\int_0^{\om_c} d \om \om^2 e^{-\om/\tau}D(\al) - \qq +
{1 \over {16 \tau^2}}\ln g \bar q \sigma G q \rn +...\label{hl}
\ee
where $\om_c$ is the duality interval, $\tau$ the Borel variable and
$D(\al)$ as defined in (\ref{ddd}), but of the form defined in (\ref{dc}).
It is, corresponding to the particular problem of heavy quarks, given as:
\be
D(\al)_B = 1 + {a_B \al}\left[1+{\lambda^2 \over {\tau^2}}
\left({\alpha(\lambda^2) \over {\alpha(\tau^2)}}\right)^{-18/11}\right]
\ee
where $a_B =   17/3+4\pi^2/9-4$log$(\om/\mu)$, with $\mu$
is chosen to be 1.3 GeV.

With the duality interval of about 1.2-1.4 GeV which is little smaller than
the
onset of QCD which corresponds to 2 GeV and $\bl \geq$ 0.6 GeV,
we get 
\be
\lambda^2 = 0.35~\mbox{GeV$^2$}. \label{l2m}
\ee

{\it Baryon}:
For the heavy baryon current
\be
j(x) = \epsilon^{abc}(\bar q_1(x)C \gamma_5 t q_2(x))Q(x)
\ee
where $C$ is charge conjugate matrix, $t$ the antisymmetric flavour matrix
and $a, b, c$ the colour indices, the QCD sum rules is given \cite{dai} by
\be
{1 \over 2} f_{\Lambda_b}^2 e^{\bl/\tau} =
{1 \over {20\pi^4}}\int_0^{\om_c} d \om \om^5
e^{-w/\tau}D(\al)_{\lbb}+
{6 \over {\pi^4}}E_G^4 \int_0^{\om_c} d \om e^{-\om/\tau}+
{6 \over {\pi^4}}E_Q^6e^{-m_0^2/8\tau^2}
\ee
where 
\be
D(\al)_{\lbb} = 1-{\al \over {4\pi}}a_{\lbb} \left(1+{\lambda^2 \over {\tau^2}}\right)
\ee
with $a_{\lbb} = r_1$log$(2\om/\mu)-r2)$. With
$f_{\Lambda_b}^2$ = $0.2 \times 10^{-3}$~GeV$^6$,
$\ln \bar q q \rn  = -0.24^3$ GeV$^3$, $\ln g \bar q \sigma G q \rn =
m_0^2 \ln \bar q q \rn$, $m_0^2 = 0.8$ GeV$^2$, $\ln \al GG \rn
= 0.04$ GeV$^4$ and $D(\al)_{\Lambda_b}$ is expressed in accordance with
power correction factor found in \cite{dai}.
As in the meson case, we obtain 
\be
\lambda^2 = 0.4~\mbox{GeV$^2$}.
\ee
Now we turn to the heavy quark expansion. The total decay
rate of a weakly decaying heavy hadron is, at the leading order, given by
\be
\Gamma(H_b) = \Gamma_0
\left[1-{\al \over \pi}\left( {2 \over 3}g(x) - \xi \right) \right] \label{rt}
\ee
where 
\be
\Gamma_0 = {G_f^2 |V_{KM}|^2 m_b^5 \over {192 \pi^3}} f(x)
\ee
As already mentioned, the power corrections are given by
the IR renormalons:
\be
D_{IR} = \tilde a \al \lln 1+ {\lambda^2 \over {m_b^2}} \lln {\alpha(\lambda^2)
\over {\alpha(m_b^2)}} \rln ^{11/18} \rln
\ee
In (\ref{rt}), the factor $\xi$ corresponds to the IR renormalons
which corresponds to the square root of the $\lambda^2$ term in the
above equation. These corrections are estimated to be 0.1$\Gamma_0$
and 0.11$\Gamma_0$
for $B$ and $\Lambda_b$ respectively. This is significant in view
of the discrepancy between the lifetimes of $B$ and $\Lambda_b$
being 0.2 ps$^{-1}$ with $\Gamma(B)$ = 0.68 ps$^{-1}$ and
$\Gamma(\Lambda_b)$ = 0.85 ps$^{-1}$.

\chapter{Quark-hadron duality}

In this Chapter, we discuss qualitatively the isssue of the quark-hadron
duality, for breveity {\it duality}. It is based on the assumption
of the convergence of the heavy quark expansion (Chapter 3) and the 
corrections arising due to renormalons contributions (Chapter 4).

\subsubsection{Prologue}
The issue of duality is as old as QCD itself. The idea of duality can be
formulated in many ways. The central idea stems from the fact that the
hadronic quantities are defined in the Minkowsky region, whereas in terms
of quarks and gluons they are specified in the Euclidean region. The two 
regions are connected by analytical continuation which is done by hand. 
However, the lore is that the hadronic observables can be calculated in 
terms of partons at a particular kinematic region. In their classic paper, 
Poggio, Quinn and Weinberg proposed a procedure, known as {\it smearing}, 
that the calculation of physical quantities can be obtained in terms of 
partons by averaging over a suitable energy range \cite{pqw}.  

In the OPE procedure, as outlined in the previous Chapter, the hadronic 
quantities are given by the leading parton quantity plus some local operators 
representing nonperturbative aspects. For example, for the leptonic decay constant 
of a meson, say $B$, we have
\be
f_B^2 \sim \int_0^{\om_0} d\om ~\rho(\om) + ~\mbox{nonperturbative corrections}
\ee
where $\rho(\om)$ represents spectral density of quark(s) and $\om$ is the duality 
interval. The determination of the duality interval depends on the contribution 
coming from the nonperturbative physics. On the other hand, the perturbative 
corrections to the leading term is also important. However, for any quark level 
quantity, the perturbative corrections are not known beyond the first few orders.

The inclusive decays of the heavy hadrons are described by the OPE based expansion 
of the hadronic matrix elements in the inverse powers of the heavy quark mass. 
The idea of quark-hadron duality is the underlying assumption. The 
discrepancy between theory and experiment for quantities given in Table \ref{t31}
casts doubt on the validity of the assumption of duality. 
According to the HQE, the total decay rate of a heavy hadron scales like the 
fifth power of the heavy quark mass concerned. As we have seen in the previous 
Chapters, the corrections to the leading order (partonic) decay rate arises
due to bound state effects which start appearing at the second order in $1/m_Q$. 
These corrections are of the order of about 10\% of the free heavy quark decay rate. 
It is significant for the HQE predictions.  

In this context, \cite{mas} have 
shown that the violation of duality in HQE is of exponential/oscillating in 
nature:
\be
\Pi(Q^2)_{violation} = e^{-C Q^2/\lq^2}
\ee
where $C$ is constant and $Q^2$ is the energy scale. However, this violating
effect is not quantified. They attribute this quantity for the discrepancy
in the inclusive properties shown in Table (\ref{t31}). 

On the other hand, in \cite{bmn}, the weak decay of heavy hadrons is studied
in the 't Hooft model. It has been found that the duality holds good with 
the presence of terms of order $1/m_Q$. Such a term is absent in the HQE. 
We should note that the first-power-suppressed term is absent in the OPE
itself. We should note that the 't Hooft QCD is 1+1 dimensional where
confinement is bulit-in. But, in QCD, the phenomenon of confinement is
not understood. Therefore, we cannot expect every aspect of the two-dimensional 
QCD to agree in {\it toto} with the QCD.

With these remarks, we discuss briefly below the qualitative aspects
derived from the study in Refs. \cite{arun2,arun}. 

\subsubsection{Perturbative corrections at large order}
The renormalon corrections evaluated in the previous Chapter for
the $B$ meson and $\lbb$ baryons are obtained in terms of the gluon mass,
$\lambda^2$ = 0.35 and 0.4 GeV$^2$ respectively \cite{arun}. The assumption on 
$\lambda^2$ is made to represent the short distance nonperturbative
effect. They correspond to the IR renormalons in the description of 
the inclusive decays:
\be
\Delta \Gamma(H_Q)_{IR} \approx a_0 \alpha_s \times \sqrt{{\lambda^2 
\over {m_Q^2}}}\left({\alpha(\lambda^2) \over {\alpha(m_Q^2)}} \right)^
{-9/11}
\ee
Numerically, the IR renormalon corrections are found to be
\bea
\Delta \Gamma (B)_{IR} \approx 0.1 \Gamma_0\\
\Delta \Gamma (\lbb)_{IR} \approx 0.11 \Gamma_0
\eea
where $\Gamma_0$ $b$-quark decay rate at the tree level. Since the sign of these
contributions is fixed to be positive, the short distance nonperturbative 
corrections yield a small but significant enhancement of the $\lbb$ decay
rate.

These corrections have to be construed as duality violating effects. 
Quantitatively, the violations amount to be about 10\%. In recent
literature, these duality violating effects  are argued to be the $1/Q^2$ 
corrections. However, it cannot have operators representations.
In the case of HQE, if it is considered to be $1/m_Q$ corrections, we
cannot have the corresponding operators.

Our conclusion is that these corrections are important in predicting
the inclusive properties in HQE.

\subsubsection{HQE for beauty mass}
In the Chapter 2, we have evaluated the EVFQO from the differences in the
inclusive decay rates \cite{arun2}. 
Assuming the convergence of the heavy quark expansion is valid one 
as long as there is an uncertainty of few percent. This would also 
positively suggest that the quark-hadron duality violating oscillating 
component might be small. So such violation would not deter a decent 
determination of quantities of interest in the heavy quark expansion
like the lifetimes of beauty hadrons, the semileptonic branching ratio
and the charm counting.
For the reasons mentioned in the beginning, the assumption on convergence 
cannot be made for charmed case. 

There are renormalon contribution from the perturbative part of the expansion.
They are IR renormalons of the order $\Lambda_{QCD}$. We, as of now, don't
have deep insight of it. They would be expected to differ for meson and
baryon. Because this contribution does not correspond to any local
operators of the theory and independent of the heavy quark mass, 
a difference of about 50 to 100 MeV between 
meson and baryon would imply much significance for the quantities
described by the heavy quark expansion. We cannot on obvious terms
argue that renormalons are related to the assumption of quark-hadron duality.
On the other hand, it will shed light on the quantities concerned and the
underlying assumption versus the heavy quark expansion. 

Our studies show that the inclusive decays of heavy hadrons can be studied within
the framework of the HQE, notwithstanding the aspects like exponential violation 
\cite{cds,bgl1} which have not been quantified.

\chapter{Conclusion}

In this Thesis, we have addressed two important issues, namely, the
evaluation of EVFQO and the quark-hadron duality in the heavy quark expansion.
Our work primarily shows that the existing discrepancy between
theory and experiment over the ratio of lifetimes of $\Lambda_b$
baryon and $B$ meson can be explained within the heavy quark expansion.
The assumption of quark-hadron duality
is found to be reasonably working well in the beauty hadrons,
since the $b$ quark
is sufficiently heavy.

The results of our study concerning the EVFQO and the ratio
of lifetimes of $\lbb$ and $B$ are given below:
\begin{itemize}
\item
The ratio of EVFQO of $\lambda_b$ and $B$ should be greater than unity,
contrary to common expectations. Precisely, it should be in the range of
three to four. Our estimation in the potential model is supported
by out another evaluation using the difference in decay rates of
$SU(3)$ triplet hadrons. The latter is r model independent result,
which depends only on the heavy quark expansion and the $SU(3)$ flavour
symmetry.
Our result for the ratio $\tau(\Lambda_b)/\tau(B)$, from the potential
model, is:
\be
\tau(\Lambda_b)/\tau(B) \approx 0.79 - 0.84
\ee
where the lower limit is fixed by the factorisable parts of the
FQO and the upper limit by the inclusion of non-factorisable
parts.
The same quantity, in the model independent approach, is
in the range of about 0.78 to 0.81.

\item
The ratio $|\Psi(0)|_{\Lambda_b}^2/|\Psi(0)|_B^2$ is about 3.8
from the potential model estimation, and about 3.5 from the model
independent approach. The latter value and the corresponding
prediction would change, if the various structures of FQO are
included. In both cases, we have estimated the $V-A$ structure
only. For the purpose calculating the ratio of lifetimes,
what we have done is sufficient. To calculate the absolute
lifetime, one has to take into account all the possible structures
of currents.

\item
The corrections due to renormalons causes an increase in the decay rate of
$\lbb$. This should also have to be taken into account in the calculation
of the inclusive properties of the heavy hadrons.

\end{itemize}
Our potential model, approach has an apparent draw back that the potential
has a term of $r^2$ term for baryon. In fact, the potential that
generates the heavy hadron mass spectrum is different. However, this potential
too satisfies the Schr\"odinger equation. We have used the same
wave function for meson and baryon.
For the latter case, it is an approximation.

Recently, in \cite{sach}, an exploratory study of the evaluation
of the EVFQO of $\lbb$ showed that the decay rate of the baryon
can be significantly enhanced due to the third order term in the
heavy quark expansion. On the other hand, in the QCD sum rules approach,
the estimated EVFQO yields the ratio of lifetimes of $\lambda_b$ and $B$ to be 0.79.

For the first time, we have studied the inclusive
charmless semileptonic decay of $\Lambda_b$. It is significant,
in view of the experimental estimate of the inclusive charmless
semileptonic branching ratio of beauty hadrons. This quantity
is useful for a clean extraction the CKM matrix element, $|V_{ub}|$.
We have observed that in measuring $Br(b \ra X_u l \nu)$, one should have to
take the spectators contribution in the baryon. Our prediction of branching ratio
including spectator effects is:
\be
Br(b \ra X_u l \nu) = 1.16 \times 10^{-3}
\ee
This result is significant in view of the recent measurement
of these quantities at LEP by ALEPH, L3 and DELPHI collaborations.

Our observations on the issue of quark-hadron duality in
the heavy quark expansion are the following:
\begin{itemize}
\item
For the $b$-quark mass being asymptotically heavy enough,
the inclusive decay rate of beauty hadrons apparently converges
at the third order in $1/m_b$. But, it is not so in the case of the
charm sector. As the results show, the EVFQO obtained for the $B$ meson and
$\lbb$ baryon yields the ratio of their lifetimes closer to the
experimental value. There is no reason to expect contribution to come
from beyond the $O(1/m_b^3)$.

\item
The implication of the assumption of the convergence of the HQE
is that quark-hadron duality works well. On the other hand, the
exponential/oscillating violation of duality
cannot be expected to be more than a few percent.

\item
Our study of the renormalons contribution to the
perturbative decay rate is important. If it is a duality breaking correction,
as we estimated, it is about 20\% of the free heavy quark decay rate.
It may be taken as the $1/m_b$ correction, but we don't have an operator
of appropriate dimension. This means that we rule out the possibility
of the absence of the $1/m_Q$ term being responsible for the duality
violation.
\end{itemize}
Thus our conclusion is that the quark-hadron duality
holds good in the HQE. The violation of few percent is within
the control.

In view of the important task of predicting the elements of the CKM matrix, the clear
idea of inclusive decays of heavy hadrons would be useful to reduce the error. Our studies
are simple demonstration concerning the two aspects of the heavy quark expansion, namely,
the systematic accommodation of the spectator effects and the knowledge of the extent to which
the notion of duality is reliable.

\end{document}